\documentstyle[epsf]{laa}
\begin{document}

\thesaurus{...}

\title{A standard stellar library for evolutionary synthesis \\I. Calibration
of theoretical spectra}

\author{Th. Lejeune \inst{1,2}
 \and    F. Cuisinier \inst{1,3} 
 \and    Roland Buser   \inst{1}}

\institute{Astronomisches Institut der Universit\"at Basel,
           Venusstr.  7,        CH--4102  Binningen,  Switzerland \and
           Observatoire de  Strasbourg,   11, rue  de  l'Universit\'e,
           F--67000 Strasbourg,  France  \and  Instituto Astronomico e
           Geofisico da   Universidade  Sao  Paulo,   Departamento  de
           Astronomia,  Caixa  Postal  9638,  01065-970 Sao Paulo  SP,
           Brasil}

\offprints{Roland Buser}
\maketitle

\begin{abstract}

We present a comprehensive hybrid library of synthetic stellar spectra
based  on three original grids of  model  atmosphere spectra by Kurucz
(1995), Fluks {\em et al.\/} (1994), and Bessell {\em et al.\/} (1989,
1991), respectively. The    combined  library has  been  intended  for
multiple-purpose synthetic photometry applications and was constructed
according to the precepts  adopted by Buser \&  Kurucz (1992):  (i) to
cover    the largest possible  ranges   in  stellar parameters  (${\rm
T_{eff}}$, log g, and [M/H]); (ii) to provide flux spectra with useful
resolution on  the  uniform  grid  of wavelengths  adopted  by  Kurucz
(1995); and  (iii)  to provide  synthetic  broad--band colors which are
highly realistic for  the largest  possible  parameter and  wavelength
ranges.

Because   the  most   astrophysically    relevant   step  consists  in
establishing a  {\em  realistic\/}  library, the corresponding   color
calibration is described in some  detail. Basically, for each value of
the  effective temperature and for each  wavelength,  we calculate the
{\em  correction function\/} that  must be applied  to a (theoretical)
solar--abundance   model  flux spectrum in   order   for this to  yield
synthetic UBVRIJHKL colors matching the (empirical) color--temperature
calibrations derived  from  observations.    In this way,     the most
important  systematic differences existing  between the original model
spectra and the  observations can indeed be  eliminated.  On the other
hand,    synthetic   UBV    and    Washington     ultraviolet excesses
$\delta_{(U-B)}$   and $\delta_{(C-M)}$  and      $\delta_{(C-T_{1})}$
obtained from  the  original  giant  and  dwarf model spectra  are  in
excellent accord with empirical metal--abundance calibrations (Lejeune
\& Buser 1996).  Therefore, the  calibration algorithm is designed  in
such  a    way as to {\em  preserve    the original  differential grid
properties implied by  metallicity and/or luminosity changes\/} in the
new library,  if the above correction   function for a solar--abundance
model  of a given effective temperature  is  also applied to models of
the same temperature  but different chemical compositions [M/H] and/or
surface gravities log g.

While  the new library constitutes  a first--order approximation to the
program set out above, it  will be allowed  to develop toward the more
ambitious goal of  matching the full  requirements  imposed on  a {\em
standard  library\/}.   Major input for  refinement  and completion is
expected from the  extensive tests  now being  made in population  and
evolutionary  synthesis  studies of the  integrated  light of globular
clusters (Lejeune 1997) and galaxies (Bruzual {\em et al.\/} 1997).

\end{abstract} 
  
\section{Introduction} 

The  success of population and  evolutionary synthesis calculations of
the integrated light of clusters  and  galaxies critically depends  on
the availability   of a suitable library   of stellar  spectral energy
distributions (SEDs), which  we  shall  henceforth call  {\em  stellar
library\/}.  Because of  the complex nature  of the subject, there are
very  many ways  in  which such   calculations  can contribute  to the
solution    of any  particular   question  relevant   to  the  stellar
populations and their evolution in  clusters and galaxies. In previous
studies,  the  particularities  of    these  questions  have   largely
determined the properties that the  corresponding stellar library must
have in  order  to be considered   {\em suitable\/}  for the  purpose.

There is now a considerable arsenal of observed stellar libraries (for
a recent compilation, see e.g. Leitherer {\em et al.\/} 1996), each of
which  has  its  particular   resolution,   coverage,  and  range   of
wavelengths as well as  its particular coverage  and range of  stellar
parameters -- but which, even if taken in the aggregate, fall short by
far of  providing   the  {\em uniform, homogeneous,   and  complete\/}
stellar library  which is required   now for  a  more systematic   and
penetrating    exploitation of   {\em   photometric\/} population  and
evolutionary synthesis.

Ultimately for this purpose, what is needed is a uniform, homogeneous,
and complete {\em theoretical\/} stellar   library, providing SEDs  in
terms of physical parameters consistent with empirical calibrations at
all accessible wavelengths. Thus, the  above goal can be approached by
merging existing grids  of  theoretical model--atmosphere spectra  into
the desired  uniform and complete stellar library,  and making it both
homogeneous and realistic by {\em empirical calibration\/}.

A variant of this approach was first tried  by Buser \& Kurucz (1992),
who constructed  a  more complete theoretical  stellar  library  for O
through K  stars by merging the  O--G--star grids of  Kurucz (1979a,b)
with the  grids of Gustafsson  {\em  et al.\/}  (1975), Bell  {\em  et
al.\/} (1976),  and Eriksson {\em et  al.\/} (1979) for F--K stars. In
their paper, Buser \& Kurucz solved for  uniformity and homogeneity by
recomputing new late--type spectra  for Kurucz's (1979)  standard grid
of  wavelengths and  using the   Kurucz \&  Peytremann (1975)   atomic
opacity  source   tables.   The resulting  hybrid library\footnote{The
Buser--Kurucz library is a {\em hybrid\/} library in the sense that it
is based on {\em two distinct grids of model atmospheres\/} which were
calculated using different codes  (ATLAS AND MARCS, respectively);  it
is  {\em quasi--homogeneous\/}, however,  because   for both grids  of
model atmospheres   the  {\em spectrum   calculations\/} were obtained
using   the  {\em  same    opacity source  tables\/}.}    has, indeed,
significantly  expanded the   ranges   of  stellar   parameters    and
wavelengths for which  {\em  synthetic photometry\/} can  be  obtained
with  useful   systematic  accuracy   and consistent   with  essential
empirical effective temperature and metallicity calibrations (Buser \&
Fenkart 1990, Buser \& Kurucz 1992, Lejeune \& Buser 1996).

In  the  meantime,  Kurucz   (1992,  1995)   has provided  a    highly
comprehensive    library of  theoretical      stellar SEDs  which   is
homogeneously  based on the  single extended grid of model atmospheres
for O to late--K stars calculated from the latest version of his ATLAS
code and using  his recent  multi--million  atomic and molecular  line
lists. The new Kurucz grid -- as we shall call it henceforth -- indeed
goes a long way toward the {\em complete library\/} matching the basic
requirements imposed by synthetic photometry studies in population and
evolutionary  synthesis. As  summarized  in  Table 1 below,  SEDs  are
provided for uniform grids of  wavelengths and stellar parameters with
almost complete coverage  of their observed  ranges !  These data have
already been widely used by  the astronomical community, and they will
doubtlessly continue to prove an indispensable database for population
and evolutionary synthesis work for years to come.

In the present  work,  we   shall endeavor   to provide yet    another
indispensable step  toward a {\em more  complete\/} stellar library by
extending  the  new Kurucz grid to   {\em cooler temperatures\/}. This
extension  is particularly important for the  synthesis of old stellar
populations, where {\em  cool giants and supergiants\/} may contribute
a  considerable fraction of the total  integrated light. Because model
atmospheres and  flux spectra for such stars   -- the M  stars -- have
been specifically calculated  by  Bessell {\em et al.\/}  (1989, 1991)
and by Fluks {\em et al.\/} (1994), our task will mainly be to combine
these with the new  Kurucz grid by transformation  to the same uniform
set of  wavelengths, and to submit the  resulting library to extensive
tests for its  realism. In fact, as  shall be shown below, the process
will  provide  a  complete grid  of  SEDs  which is  homogeneously and
consistently {\em  calibrated  against  observed  colors\/}  at   most
accessible wavelengths.

In Sect. 2, we shall briefly describe  the different libraries used in
this paper and   the   main  problems   that   they   pose to    their
unification. Because  the spectra exhibit  systematic differences both
between  their parent libraries  and  relative to observations, we set
up, in Sect. 3, the basic empirical color--temperature relations to be
used for uniformly calibrating the library  spectra in a wide range of
broad--band colors. This calibration process  is driven by a  computer
algorithm  developed  and   described   in    Sect. 4.  The    actual
color--temperature relations  obtained   from  the  corrected  library
spectra are discussed  in Sect. 5, and the  final  organization of the
library grid is  presented in Sect.  6. In the  concluding Sect. 7, we
summarize  the present state of   this work toward  the intended  {\em
standard library\/},   and we briefly   mention those  necessary steps
which are currently in process to this end.

\section{The basic stellar libraries}

The different libraries used  are from Kurucz  (1995), Bessell {\em et
al.\/} (1989, 1991), and Fluks {\em et al.\/} (1994) -- which we shall
henceforth call the K--, B--, and F--libraries, respectively. Although
the   K--library covers a   very  wide temperature  range (from  ${\rm
T_{eff}}$=50,000  K to  3500 K),  it does not  extend  to the very low
temperatures required  to model cool AGB  stars. These stars  are very
important for population  and  evolutionary synthesis, since they  can
represent up  to 40\% of  the  bolometric and even up   to 60\% of the
K--band luminosity of a single stellar  generation (Bruzual \& Charlot
1993).   It is natural, then,  to  provide the necessary supplement by
employing the suitable libraries that were specifically calculated for
M--giant stars in the  temperature range 3800--2500  K by Bessell {\em
et al.\/} (1989,1991) and by Fluks {\em et al.\/} (1994).

Table 1 summarizes the coverage of parameters and wavelengths 
provided by these three libraries, and Figs. 1 and 2 illustrate original 
sample spectra as functions of metallicity for two temperatures. 

\begin{table*}
\caption []{Parameter and wavelength coverage of the different libraries}
\begin{tabular}{cccc}
\hline
    & Kurucz (1995) & Bessell {\em et al.} (1989,1991) & Fluks {\em et
    al.} (1994)\\
\hline\hline
${\rm T_{eff}}$ & 3500 $\sim$ 50,000 K & 2500 $\sim$ 3800 K & 
                                                      2500 $\sim$ 3800 K \\
log g   & 0.0 $\sim$ 5.0 & -1.0 $\sim$ +1.0  & red giant sequence \\
${\rm [M/H]}$   & -5.0 $\sim$ +1.0 & -1.0 $\sim$ +0.5 & solar \\
$\lambda\lambda$(nm) & 9.1 $\sim$ 160,000 & 491 $\sim$ 4090 & 
                                                      99 $\sim$ 12,500 \\
${\rm n(bin_{\lambda})}$ & 1221 & 705 & 10,912 \\
\hline
\end{tabular}
\label {t:param}
\end{table*}

\begin{figure}
\epsfxsize=9cm
\epsffile {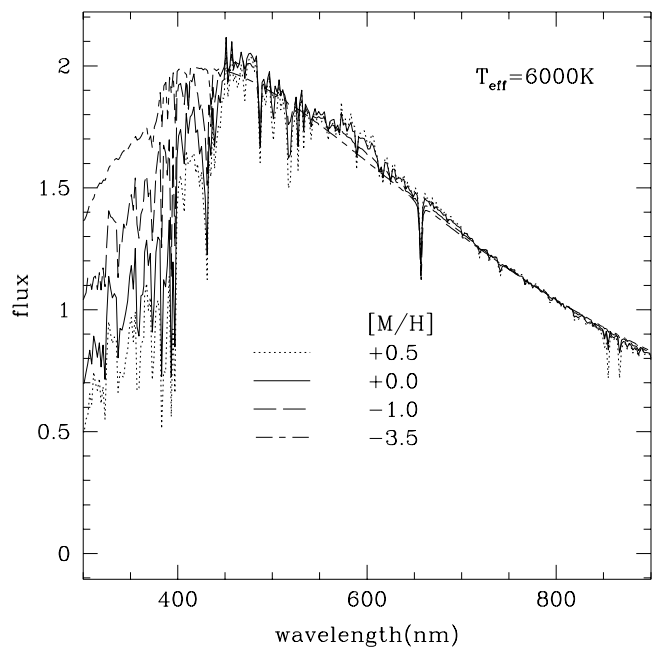}
\caption [] {K--library dwarf model spectra $(log g = 5)$ for a range in
        metallicity. All the spectra are normalized at $\lambda = 817 nm$.}
\label{f:sp6000}
\end{figure}

\begin{figure}
\epsfxsize=9cm
\epsffile {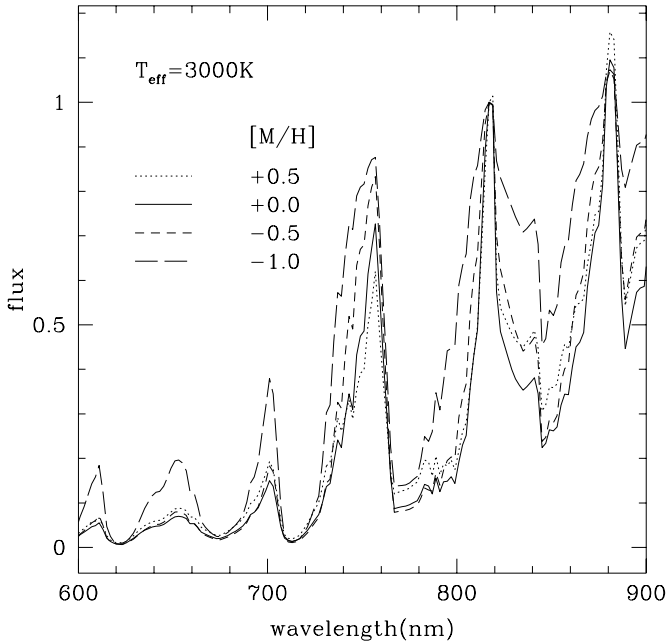}
\caption [] {B--library giant model spectra $(log g = -0.29)$ for a range
        in metallicity. All the spectra are normalized at $\lambda = 817 nm$.}
\label {f:sp3000}
\end{figure}

We should note  the  following: first,  while both   the K-- and   the
B--library spectra are given  for different but overlapping  ranges in
metallicity,  the F--library spectra  have   been available for  solar
abundances only; secondly,   the B--spectra are given for  wavelengths
$\lambda\lambda\geq491nm$       and  nonuniform   sampling,  while the
F--spectra   also       cover     the    ultraviolet       wavelengths
$\lambda\lambda\geq99nm$ in a uniform manner.

As it is  one of the goals  of  the present effort to  eventually also
allow  the  synthesis  of the metallicity--sensitive near--ultraviolet
colors  (e.g., the Johnson U and  B), the  B--library spectra -- which
account for the spectral  changes due to  variations in metallicity --
have been   complemented by  the  F--library spectra   at  ultraviolet
wavelengths.

This step was achieved with the following procedure:

\begin{description}
\item[--]
All the F--  and B--spectra  have first  been resampled at  the (same)
uniform grid of wavelengths given by the K--library spectra.
\item[--]
F--spectra were then   recomputed   for the   effective   temperatures
associated with the B--spectra. This was  done by interpolation of the
F--library sequence of 11  spectra representing M--giants of  types M0
through    M10, and using  the spectral   type -- $\rm T_{eff}$ scale
defined by Fluks {\em et al.\/}.
\item[--]
Finally, in  order  to avoid  the spurious  spikes  at $\lambda\lambda
510nm$   present in  the  synthetic  B--spectra  (Worthey  1994), {\em
each\/} B--spectrum was  combined with  the blue  part ($9.9  \sim 600
nm$) of  the  corresponding F--spectrum, i.e.,   having the same  $\rm
T_{eff}$ and  having been rescaled  to  the B--spectrum flux  level at
$\lambda\lambda 600 nm$.
\end{description}

The hybrid  spectra created  in  this manner  will hereafter be called
`B+F--spectra'.

Of course in this way, completeness in wavelength coverage can only be
established  for  solar--abundance models. However,  extensions of the
B--spectra to optical  wavelengths down to   the atmospheric limit  at
$\lambda =   320nm$ are being  worked out  now and will  supersede the
corresponding  preliminary hybrid B+F--spectra  (Buser {\em et al.\/}
1997).

\section{Comparison with empirical calibrations}

In  order to assess  the reliability of the  synthetic spectra, we now
compare them to  empirical temperature--color calibrations.  Depending
on the  availability and the  quality  of calibration data, two  basic
calibration sequences  will be used for the  cooler giants and for the
hotter main sequence stars, respectively.

\subsection{Empirical color--temperature relations}

\subsubsection{Red giants and supergiants}

Ridgway {\em et al.\/} (1980) derived an empirical temperature--(V--K)
relation for cool giant  stars.  This relation   is based on   stellar
diameter and flux measurements, and therefore on the definition of the
effective temperature:

\begin{equation}
f_{bol} = \left( \frac{\phi}{2} \right) ^{2}\sigma T_{eff}^{4},
\end{equation}
where  $f_{bol}$   is  the  apparent   bolometric   flux, $\sigma$  is
Boltzmann's  constant, and  $\phi$ is   the  angular diameter.   Hence
$f_{bol}$ is almost entirely empirical.

Over the range ${\rm 5000  K \sim 3250 K}$,  the Ridgway {\em et  al.}
calibration  was  adopted as the effective  temperature  scale for the
V--K colors.  We derived  the  color-temperature relations for   V--I,
J--H,  H--K, J--K and  K--L using the  infrared two-color calibrations
given by Bessell
\& Brett (1988) (hereafter referred to as BB88).   For the U--B and B--V
colors, we  used  the  color--color relations  established   by Johnson
(1966),  and Bessell's (1979)    calibration   was adopted  for    the
(R--I)--${\rm T_{eff}}$ relation.

Because existing calibrations do not go below  ${\rm \sim 3200 K}$, we
have used both observations and theoretical results published by Fluks
{\em  et  al.\/} (1994)  in  order to  construct semi--empirical ${\rm
T_{eff}-color}$ calibrations down to the  range ${\rm 3250 K \sim 2500
K}$. Synthetic V--K colors computed from their sequence of photospheric
model spectra provide a very good match  to the calibration by Ridgway
{\em et al.}, which could thus be extended to  the range (${\rm 3767 K
\sim  2500 K}$)  by  adopting  the theoretical  ${\rm  T_{eff}}$--(V--K)
relation  from   Fluks  {\em et   al.\/}\footnote{The  F-- models were
preferred  to the B-- models in  establishing the semi-empirical ${\rm
T_{eff}}$--(V--K) relation  since they    include  a  more    accurate
calculation of  the opacity (the  Opacity  Sampling technique was used
instead of the simple Straight Mean method  which causes problems when
lines saturate).   These models also  incorporate atomic data, as well
as new opacities  for the molecules  ({\em e.g.}  ${\rm H_{2}O}$, TiO,
VO), hence providing  significant a improvement  in the V--K synthetic
colors at solar metallicity (Plez {\em et al.}   1992).}  We then used
Fluks    {\em     et       al.'s\/}      compilation  of         ${\rm
(UB)_j(VRI)_c(JHKLM)_{ESO}}$ observations  of a large sample of bright
M--giant stars  in   the solar  neighbourhood  to  establish the  mean
intrinsic colors and    standard deviations from   their  estimates of
interstellar extinction  within each photometric band.   Finally, with
these  results, we derived  mean  intrinsic color--color relations  --
(V--I)-(V--K), (U--B)-(V--I),  (R--I)-(V--K)    and (B--V)-(R--I)  --,
which allowed  us  to translate to all   these other colors  the basic
${\rm  T_{eff}}$--(V--K) relation adopted above  for red giants within
the temperature range ${\rm 3250 K
\sim 2500 K}$.

Examples  of the     adopted fits  for  the   (V--I)-(V--K)   and  the
(U--B)-(V--I) sequences are shown in Figs. 3 and 4.
\begin{figure}
\epsfxsize=9cm
\epsffile{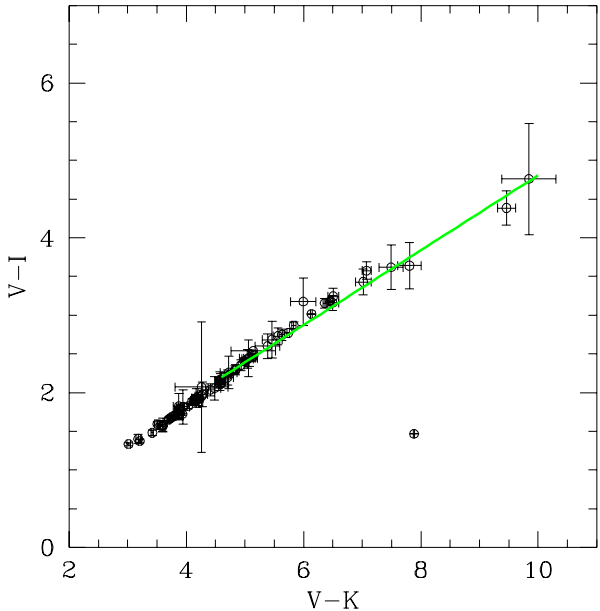}
\caption [] {Adopted (V--I)-(V--K) two--color calibration sequence for cool
        giants.  Symbols   represent    mean    values   derived  from
        observations given  by Fluks {\em et  al.\/} (1994). The solid
        line is a linear least--squares fit to the data for V-K $\geq$
        5.}
\label{vivk:fit}
\end{figure}
\begin{figure}
\epsfxsize=9cm
\epsffile{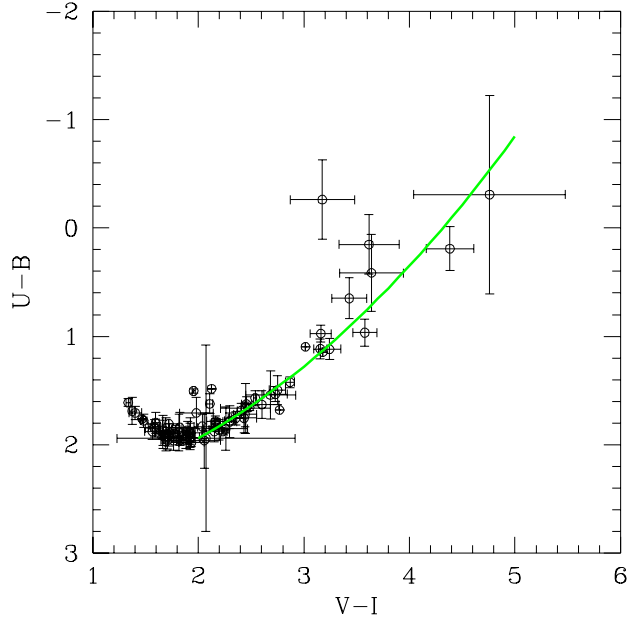}
\caption [] {Adopted (U--B)-(V--I) two--color calibration sequence for cool
        giants. Symbols  represent the  same as  in  Fig. 3. The solid
        line is a quadratic fit to the data for V-I $\geq$ 2.}
\label{ubvi:fit}
\end{figure}

For the infrared  colors, the photometric data given  by Fluks {\em et
al.}   are defined on  the   $(JHKL)_{ESO}$  filter system, which   is
different from the  filter system  defined  by BB88.  Using the  color
equations  relating the two systems and  derived  by BB88, transformed
JHKL  colors from the  Fluks  {\em et  al.}   data were computed.  The
resulting (V--K)-(V--J),        (V--K)-(V--H),       (V--L)-(V--J) and
(V--H)-(V--J) sequences are well approximated by linear extrapolations
of the   two--color relations given by  BB88.   Furthermore, the model
colors derived from the Fluks {\em et  al.}  synthetic spectra and the
filter responses defined  by BB88  also   agree very well  with  these
extrapolated relations (Fig.  5).   We therefore chose this method  to
derive the J--H, H--K, J--K and K--L colors  over the range ${\rm 3250
K
\sim 2500 K}$.  However, the uncertainty  implied by the extrapolation
to  the reddest giants  is of the order  of ${\rm 0.1  \sim 0.2}$ mag,
indicating that for the coolest temperatures near 2500 K the resulting
empirical calibration of the J--H, H--K,  J--K, and K--L colors should
be improved by future observations.

\begin{figure}
\epsfxsize=9cm
\epsffile{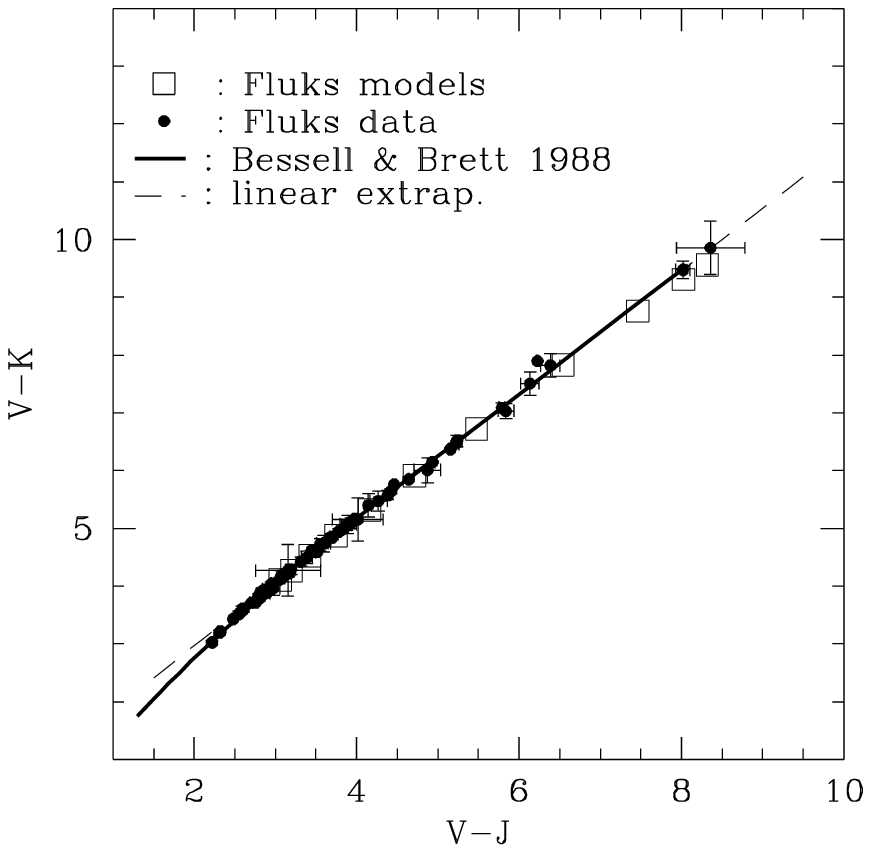}
\caption [] {Adopted extrapolation of the (V--K)-(V--J) relation for cool giants (dashed line). Model  colors (open squares) and  observed data  (crosses) from Fluks   et   al.  (1994)  are   compared  to  the  empirical  sequence adopted. (See text for explanations.)}
\label{vkvj:fit}
\end{figure}

Table 2 presents the  final adopted temperature--color calibrations for
red giants, which was supplemented by a log g  sequence related to the
effective temperature via an evolutionary  track (${\rm 1 M_{\odot}}$)
calculated by Schaller {\em et al.\/} (1992).

\begin{table*} 

\caption{Adopted semi--empirical ${\rm T_{eff}}$--color calibrations and log g values
for cool giants. (The notes refer to the method employed to derive the
semi--empirical colors at a given temperature.)}

\begin{tabular}{ccccccccccc} 
\hline 
${\rm T_{eff}}$  &  U--B     &  B--V     & V--I     &  V--K     &  R--I   & J--H     & 
 H--K     &  J--K     &  K--L   & log g \\
\hline 
\hline 
4593       & $ 1.017^{5}$ & $ 1.094^{5}$ & $ 1.080^{1}$ & $ 2.500^{1}$ & 
$0.487^{6}$ & $ 0.580^{1}$ & $ 0.100^{1}$ & $ 0.680^{1}$ & $ 0.080^{1}$ & 2.85
\\ 
4436       & $ 1.187^{5}$ & $ 1.173^{5}$ & $ 1.170^{1}$ & $ 2.700^{1}$ & 
$0.530^{6}$ & $ 0.630^{1}$ & $ 0.115^{1}$ & $ 0.740^{1}$ & $ 0.090^{1}$ & 2.50
\\ 
4245       & $ 1.399^{5}$ & $ 1.281^{5}$ & $ 1.360^{1}$ & $ 3.000^{1}$ & 
$0.602^{6}$ & $ 0.680^{1}$ & $ 0.140^{1}$ & $ 0.820^{1}$ & $ 0.100^{1}$ & 2.12
\\ 
4095       & $ 1.566^{5}$ & $ 1.364^{5}$ & $ 1.479^{1}$ & $ 3.260^{1}$ & 
$0.672^{6}$ & $ 0.730^{1}$ & $ 0.150^{1}$ & $ 0.880^{1}$ & $ 0.110^{1}$ & 1.82
\\ 
3954       & $ 1.714^{5}$ & $ 1.443^{5}$ & $ 1.634^{1}$ & $ 3.600^{1}$ & 
$0.773^{6}$ & $ 0.790^{1}$ & $ 0.165^{1}$ & $ 0.950^{1}$ & $ 0.120^{1}$ & 1.55
\\ 
3870       & $ 1.784^{5}$ & $ 1.489^{5}$ & $ 1.768^{1}$ & $ 3.850^{1}$ & 
$0.859^{6}$ & $ 0.830^{1}$ & $ 0.190^{1}$ & $ 1.010^{1}$ & $ 0.120^{1}$ & 1.39
\\ 
3801       & $ 1.815^{5}$ & $ 1.524^{5}$ & $ 1.899^{1}$ & $ 4.050^{1}$ & 
$0.948^{6}$ & $ 0.850^{1}$ & $ 0.205^{1}$ & $ 1.050^{1}$ & $ 0.130^{1}$ & 1.25
\\ 
3730       & $ 1.812^{5}$ & $ 1.552^{5}$ & $ 2.053^{1}$ & $ 4.300^{1}$ & 
$1.058^{6}$ & $ 0.870^{1}$ & $ 0.215^{1}$ & $ 1.080^{1}$ & $ 0.150^{1}$ & 1.15
\\ 
3640       & $ 1.750^{5}$ & $ 1.577^{5}$ & $ 2.269^{1}$ & $ 4.640^{1}$ & 
$1.228^{6}$ & $ 0.900^{1}$ & $ 0.235^{1}$ & $ 1.130^{1}$ & $ 0.170^{1}$ & 0.98
\\ 
3560       & $ 1.651^{5}$ & $ 1.590^{5}$ & $ 2.472^{1}$ & $ 5.100^{1}$ & 
$1.568^{4}$ & $ 0.930^{1}$ & $ 0.245^{1}$ & $ 1.170^{1}$ & $ 0.180^{1}$ & 0.83
\\ 
3420       & $ 1.412^{5}$ & $ 1.589^{5}$ & $ 2.828^{1}$ & $ 5.960^{1}$ & 
$1.899^{4}$ & $ 0.950^{1}$ & $ 0.285^{1}$ & $ 1.230^{1}$ & $ 0.200^{1}$ & 0.56
\\ 
3250       & $ 1.019^{4}$ & $ 1.527^{4}$ & $ 3.309^{4}$ & $ 6.840^{1}$ & 
$2.170^{4}$ & $ 0.960^{1}$ & $ 0.300^{1}$ & $ 1.260^{1}$ & $ 0.256^{1}$ & 0.21
\\ 
3126       & $ 0.645^{4}$ & $ 1.499^{4}$ & $ 3.709^{4}$ & $ 7.830^{3}$ & 
$2.391^{4}$ & $ 0.950^{2}$ & $ 0.370^{2}$ & $ 1.320^{2}$ & $ 0.310^{2}$ &-0.05
\\ 
2890       & $ 0.096^{4}$ & $ 1.512^{4}$ & $ 4.234^{4}$ & $ 8.760^{3}$ & 
$2.519^{4}$ & $ 0.920^{2}$ & $ 0.400^{2}$ & $ 1.320^{2}$ & $ 0.420^{2}$ &-0.57
\\ 
2667       & $-0.146^{4}$ & $ 1.507^{4}$ & $ 4.439^{4}$ & $ 9.310^{3}$ & 
$2.558^{4}$ & $ 0.900^{2}$ & $ 0.410^{2}$ & $ 1.310^{2}$ & $ 0.510^{2}$ &-1.09
\\ 
2500       & $-0.328^{4}$ & $ 1.510^{4}$ & $ 4.593^{4}$ & $ 9.560^{3}$ & 
$2.567^{4}$ & $ 0.880^{2}$ & $ 0.420^{2}$ & $ 1.300^{2}$ & $ 0.550^{2}$ &-1.52
\\ 
\hline
 & & & & & & & & & & 
\end{tabular}

\scriptsize
$^{1}$ Bessell \& Brett (1988) two--color relation with Ridgway {\em et
al.}  (1980) to relate ${\rm T_{eff}}$ to V--K.

$^{2}$ Extrapolation of two--color relations from Bessell \& Brett (1988).

$^{3}$ Synthetic color indices from Fluks {\em et al}. (1994) models.

$^{4}$ From  mean two--color relations derived from  the Fluks  {\em et
al.} (1994) observed data.

$^{5}$ Empirical calibration from Johnson (1966).

$^{6}$ Empirical calibration from Bessell (1979).

\label{empcal_gi}
\end{table*}

\subsubsection{Main sequence stars}

To construct empirical ${\rm T_{eff}}$--color sequences from 12000 K to
3600  K for the main  sequence  stars, we used different calibrations:
Schmidt-Kaler (1982) was chosen to  relate ${\rm T_{eff}}$ to U--B, B--V
or R--I, and the two--color relations  established by FitzGerald (1970),
Bessell (1979), and   BB88 were then   used to derive  the temperature
scales for  the   remaining  colors.  This  procedure   should provide
color--temperature  calibrations with uncertainties  of $\leq$ 0.05 mag
in color or $\leq$ 200 K in temperature (Buser \& Kurucz 1992).

\subsection{Comparison of theoretical and empirical color--temperature 
relations\/}

\subsubsection{Red giants and supergiants}

In order  to compare    the  models to the  above   color--temperature
relations for red giants, model spectra were first interpolated in the
theoretical libraries for  appropriate values of surface gravity given
by the  log  g--${\rm T_{eff}}$    relation defined  by the  ${\rm   1
M_{\odot}}$  evolutionary track     from  Schaller {\em   et    al.\/}
(1992).  Synthetic colors computed from  these  model spectra are then
directly compared to the empirical color--temperature relations, as
illustrated in Fig. 6.

\begin{figure*}
\epsfxsize=18cm
\epsffile{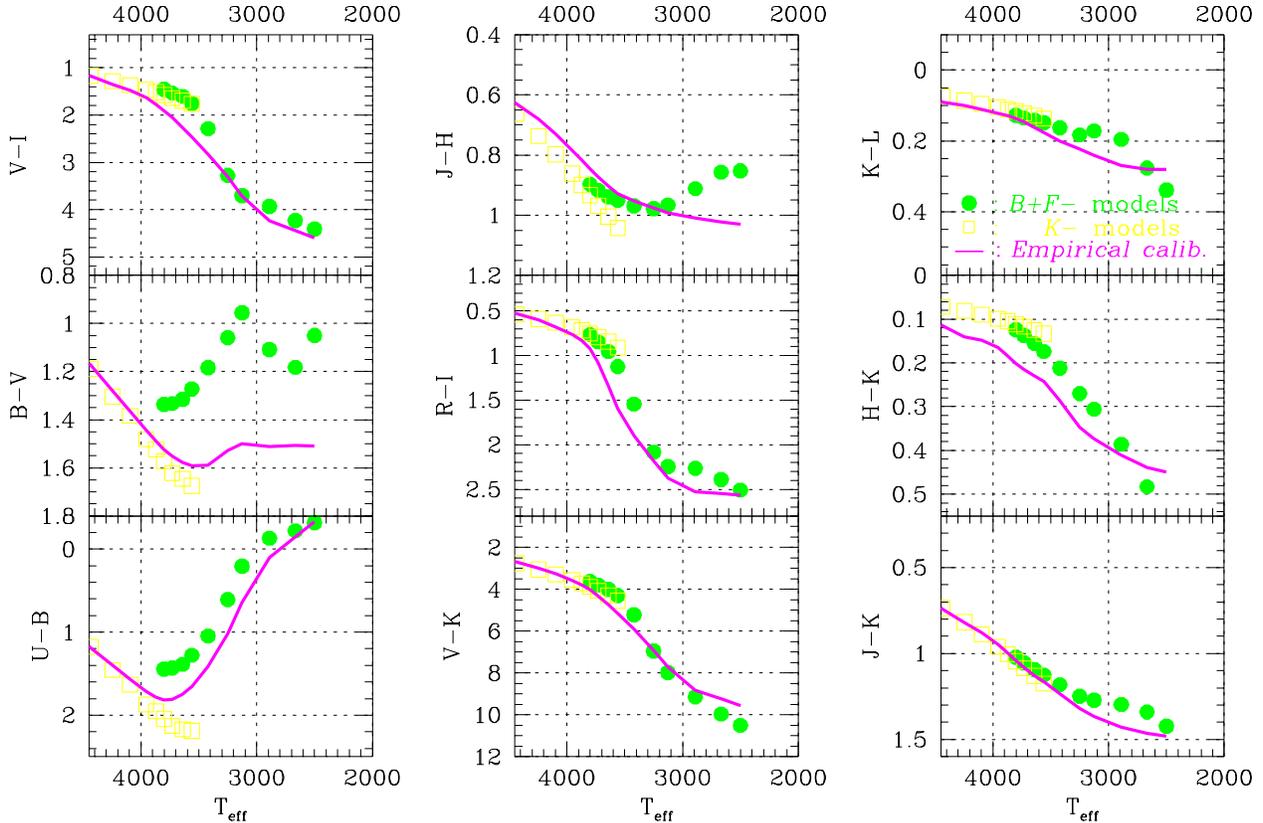} 
\caption[]{Empirical color--effective temperature calibrations for
        solar--metallicity red giant stars  (solid lines, according to
        Table 2) compared to  the corresponding theoretical  relations
        calculated  from  original synthetic library spectra (symbols,
        according to key in insert).  Note  that different scales have
        been used for the different colors.}
\label{f:gicol_uncor}
\end{figure*}

It is evident that  the  color differences between equivalent   models
from the K-- and the B+F--libraries can  be as high  as {\em 0.4 mag},
while those between the theoretical library  spectra and the empirical
calibrations may be even larger, up to {\em 1 mag}.

Such  differences -- both between   the original libraries and between
these and the empirical calibrations -- make  it clear that direct use
of these original   theoretical  data in population  and  evolutionary
synthesis is  bound  to generate a   great deal  of confusion   in the
interpretation   of results.  In    particular, applications  to   the
integrated light of galaxies at faint  magnitude levels, where effects
of  cosmological redshift may come  into  play  as well, will  provide
rather limited physical insight   unless the basic building blocks  of
the evolutionary  synthesis   -- i.e.,  the   stellar spectra --   are
systematically   consistent  with   the best   available observational
evidence.  Thus, our work is driven by the {\em systematic consistency
of    theoretical stellar colors  and  empirical  calibrations\/} as a
minimum  requirement for the (future) {\em  standard  library\/}. As a
viable operational step in this direction,  a suitable {\em correction
procedure for  the theoretical spectra\/}   will be developed in  the
following section.

\subsubsection{Main sequence stars}

The same procedure as for the giants was applied for the main sequence
stars, except that a zero--age main sequence isochrone (ZAMS) compiled
by  Bruzual (1995) was  used in the  appropriate interpolation for the
surface gravity log  g.  Again, synthetic photometry  results obtained
from the theoretical library   (Kurucz) are compared to the  empirical
color--temperature relations in Fig. 7.

\begin{figure*}
\epsfxsize=18cm
\epsffile{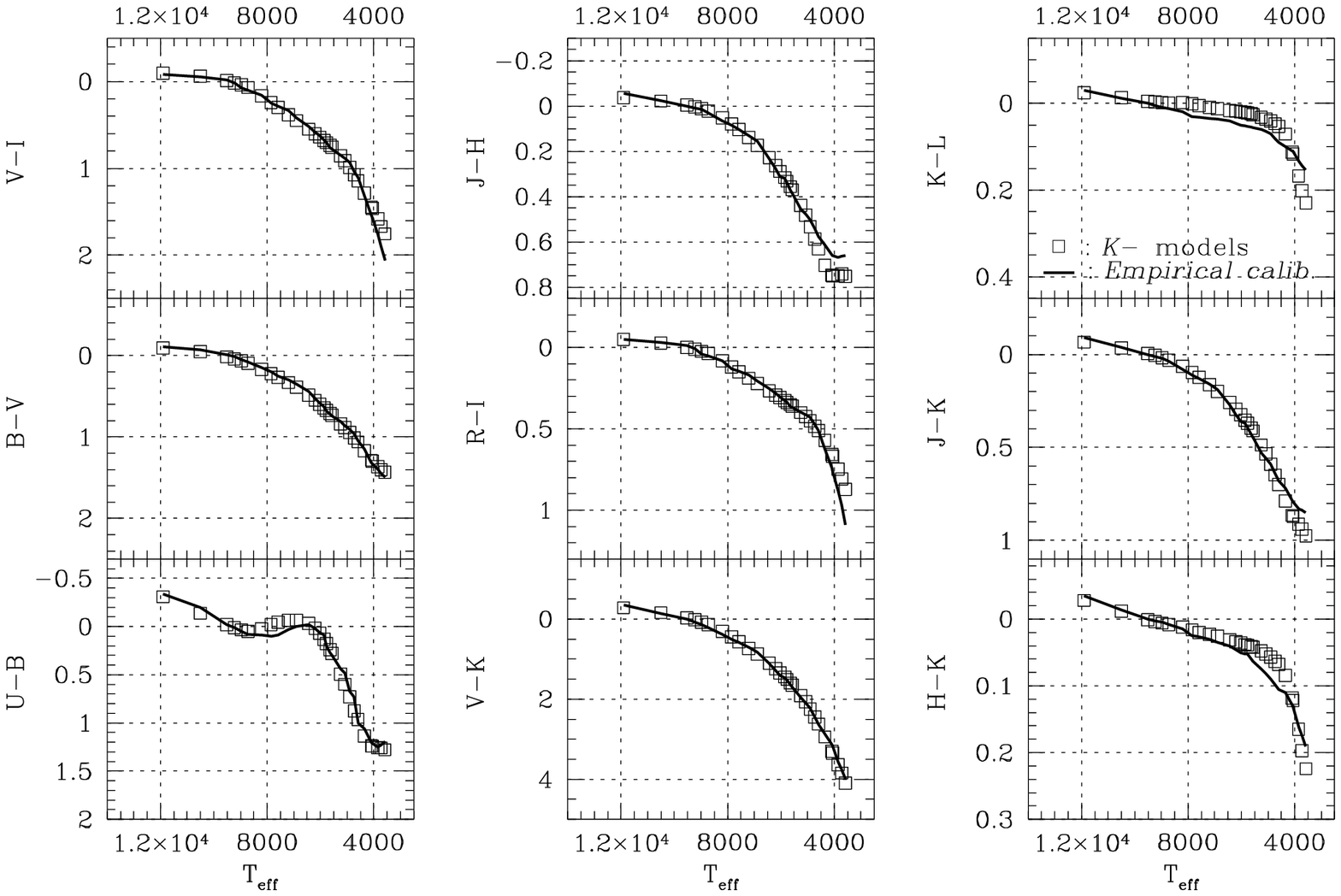}
\caption[]{Empirical color--effective temperature calibrations for
        solar--metallicity  dwarf  stars  (solid lines,   see text for
        sources) compared to  the corresponding  theoretical relations
        calculated from  original   synthetic   K--library     spectra
        (symbols).  Note that different scales have  been used for the
        different colors.}
\label{f:dwcol_uncor}
\end{figure*} 

Note that the  differences  between the theoretical and  the empirical
colors  are significantly smaller than  those  for the giants given in
Fig.  6: for  the  hotter temperatures, they  do not  exceed 0.1 mag.,
while at cooler temperatures ($\simeq 3500$ K) differences of up to 0.3
mag in V--I between models   and observations again indicate that  the
coolest K--library spectra still carry large uncertainties and should,
therefore,  be used with caution (e.g.,   Buser \& Kurucz 1992). Thus,
application of the same correction  procedure as for the giant  models
appears warranted for the dwarf models as well.

Also note that the  Kurucz spectra only go  down to 3500K. We are thus
missing the low--luminosity, low--temperature main sequence M stars in
the present library. However,  we anticipate here  that in a corollary
paper (Lejeune {\em et al.\/}  1997, hereafter Paper II) the necessary
extension  is being     provided from  a  similar   treatment  of  the
comprehensive grid of  M--star  model spectra  published by Allard  \&
Hauschildt (1995).

\section{Calibration algorithm for theoretical library spectra}

We now establish a correction  procedure {\em for the library spectra}
which preserves their detailed features but modifies their continua in
such a  way   that the synthetic   colors from  the corrected  library
spectra conform  to the empirical  color--temperature relations. Since
the empirical  color--temperature   relations do not  provide   direct
access  to the stellar continua,   pseudo--continua are instead  being
calculated for each temperature from both the  empirical (Table 2) and
the  theoretical (model--generated) colors. The  ratio between the two
pseudo--continua then provides the desired {\em correction function\/}
for the given ${\rm T_{eff}}$.

\subsection{Pseudo--continuum definition}

For a given stellar  flux spectrum of effective temperature $T_{eff}$,
we define the pseudo--continua $pc_{\lambda}(T_{eff})$ as black bodies
of color temperature $T_c(\lambda)$ varying with wavelength:

\begin{equation}
pc_{\lambda}(T_{eff}) = \alpha(T_{eff}) \cdot B_{\lambda}(T_c(\lambda)), 
\end{equation}

where $\alpha(T_{eff})$ is a scale factor and $B_{\lambda}(T)$ is the black 
body function, both of which need to be determined by least--squares fits
of $pc_{\lambda}(T_{eff})$ to the broad--band fluxes of the given flux 
spectrum. Thus, 

\begin{equation}
\alpha(T_{eff}) \cdot \int B_{\lambda}(T_{mean})S_i(\lambda) d\lambda 
\simeq f_i, i=1,...,9,
\end{equation}

where $S_i$  is the normalized transmission  function  of the passband
$i$ and $f_i$ is the  broad--band flux measured through this passband.
Because colors are relative measurements,  we normalize by arbitrarily
setting  the absolute   flux  in the  I--band  to  be  equal  to  100:
$f_{I}=100.$  The black--body fit   in    eqn. (3) is then    obtained
iteratively by a conjugate gradients method.

Fig. 8  illustrates a typical   result. Note that, because effects  of
blanketing   are    ignored  by  this   fitting    procedure, the mean
temperature, $T_{mean}$, associated with the best--fitting black--body
curve, is systematically lower  than the effective temperature of  the
actual flux spectrum.\\

\begin{figure}
\epsfxsize=9cm
\epsffile{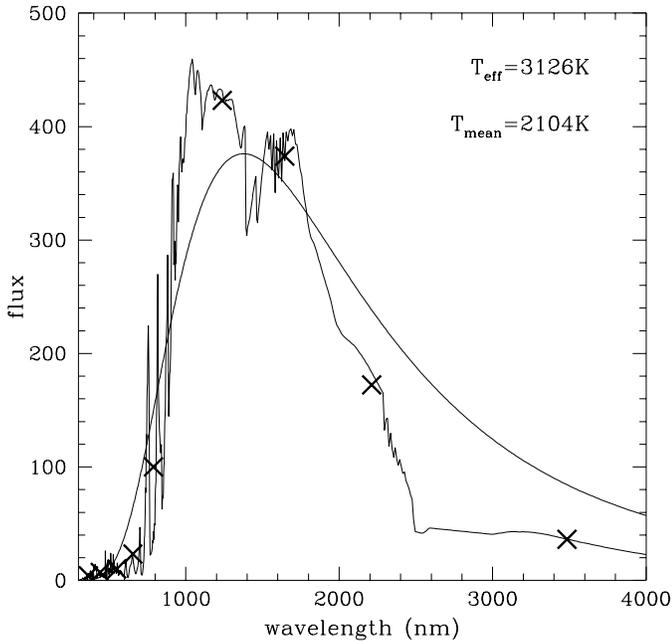}
\caption{Black--body fit of temperature
        $T_{mean}$  to   a  synthetic  spectrum  for  $T_{eff}=3126K$,
        illustrating the least--squares solution of equations (3). The
        crosses represent  the integrated  (heterochromatic) fluxes of
        the   synthetic  spectrum measured   in  the different (broad)
        bandpasses.}
\label {f:bb}
\end{figure}

$\alpha(T_{eff})$ having thus been determined, the color temperatures, 
$T_c(\lambda_i)$, can be derived in a straightforward manner at the mean 
wavelengths  
\footnote{
The mean wavelength $\lambda_i$ of a filter of transmission function
$S_i$ is defined in the following way:
\[
\lambda_i = \frac{\int \lambda S_i(\lambda) d\lambda}
                 {\int S_i(\lambda) d\lambda}.
\]}
$\lambda_i$ of the
passbands $i$ via the equations:

\begin{equation}
\alpha(T_{eff}) \cdot \int  B_{\lambda}(T_c(\lambda_i))S_i(\lambda)d\lambda 
= f_i, i=1,...,9.
\end{equation}

Interpolation  between the $\lambda_i$   by a spline function  finally
provides the continuous (and smooth) color temperatures $T_c(\lambda)$
(Fig. 10) required    to  calculate the pseudo--continua defined    by
equation (2).\\

\subsection{Correction procedure}

The   correction procedure is defined  by  the following sequence, and
illustrated (steps 1 to 4) in Fig. 10.

\begin{enumerate}

\item 
At effective  temperature  $T_{eff}$, the empirical  pseudo--continuum
$pc_{\lambda}^{emp}(T_{eff})$  is  computed from   the colors   of the
empirical temperature--color relations given in Table 2.

\item
At   the  same   effective   temperature   $T_{eff}$,  the   synthetic
pseudo--continuum  is computed from the  synthetic colors obtained for
the original theoretical   solar--abundance spectra given  in the  K--
and/or B+F--libraries:

\begin{equation}
pc_{\lambda}^{syn}(T_{eff})=  pc_{\lambda}^{syn}(T_{eff}, log g    \in
sequence, [M/H]=0),
\end{equation}

where log  g   is defined in  the    same way  as  in  Sect.  3 by  a
$1M_{\odot}$    evolutionary  track calculated   by Schaller   {\em et
al.\/} (1992).

\item
The correction function is calculated  as the  ratio of the  empirical
pseudo--continuum and the synthetic pseudo--continuum at the effective
temperature $T_{eff}$:\footnote{Note   that  in  some equations   that
follow we use a more compact notation to designate fundamental stellar
parameters which is related  to the usual  notation by the equivalence
$(T,logg,\chi)\equiv(T_{eff},logg,[M/H])$.}

\begin{equation}
\Phi_{\lambda}(T) = 
\frac{pc_{\lambda}^{emp}(T_{eff})}{pc_{\lambda}^{syn}(T_{eff})}.
\end{equation}

\item
Corrected spectra,  $f_{\lambda, corr}^{*}$, are then  calculated from
the original library spectra, $f_{\lambda}$:

\begin{equation}
f_{\lambda, corr}^{*}(T,logg,\chi)=
f_{\lambda}(T,logg,\chi)\cdot \Phi_{\lambda}(T).
\end{equation}

Note    that the correction defined  in   this way becomes an additive
constant  on  a logarithmic, or   magnitude  scale. Therefore, at each
effective temperature the original monochromatic magnitude differences
between  model library   spectra having  different metallicities [M/H]
and/or  different  surface   gravities   log g are   conserved   after
correction. As    we  shall see  below,   this will  be  true  to good
approximation even for the  heterochromatic broad--band magnitudes and
colors because,  as  shown   in Fig. 12,   the   wavelength--dependent
correction functions do not, in general, exhibit dramatically changing
amplitudes within the passbands.

Fig.  11 shows  how  the  resulting correction functions  change  with
decreasing effective temperature. Finally,  Figs.   12 and 13  display
the   corresponding  effective  temperature sequence  of  original and
corrected spectra at different metallicities   for the full (Fig.  12)
and the visible (Fig. 13) wavelength ranges, respectively.

\begin{figure}[h]
\epsfxsize=9cm
\epsffile{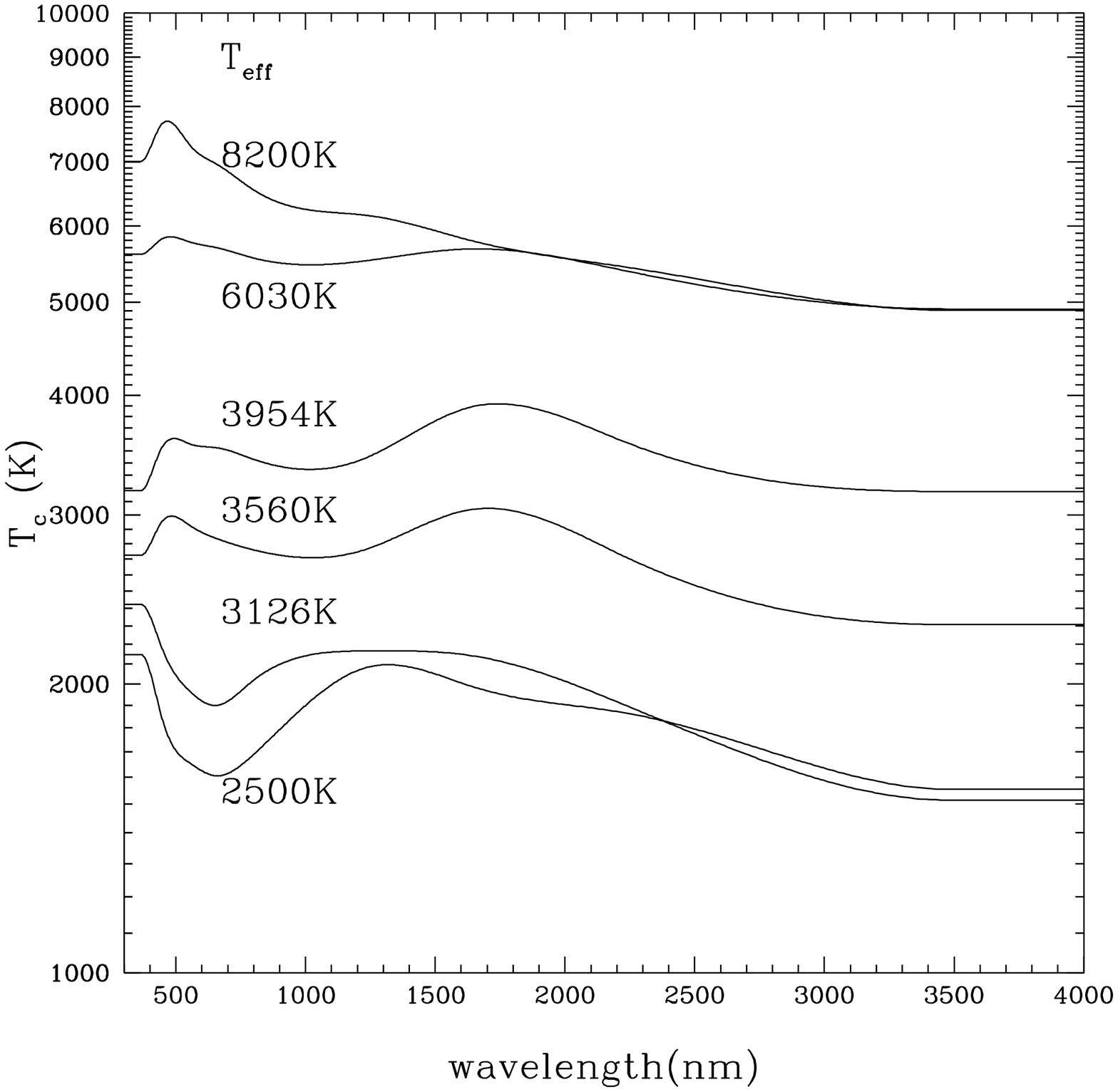}
\caption[]{Color temperatures for synthetic spectra covering a range of
        effective   temperatures,  as  labelled. Top  pair: K--library
        dwarf  models;  middle  pair:  K--library giant models; bottom
        pair:   B+F--library giant models.   Note  that systematically
        $T_{c}  < T_{eff}$, as  tracked down  by  the black--body  fit
        temperatures, $T_{mean}$.}
\label{f:tc}
\end{figure}

\begin{figure*}
\epsfxsize=18cm
\epsffile{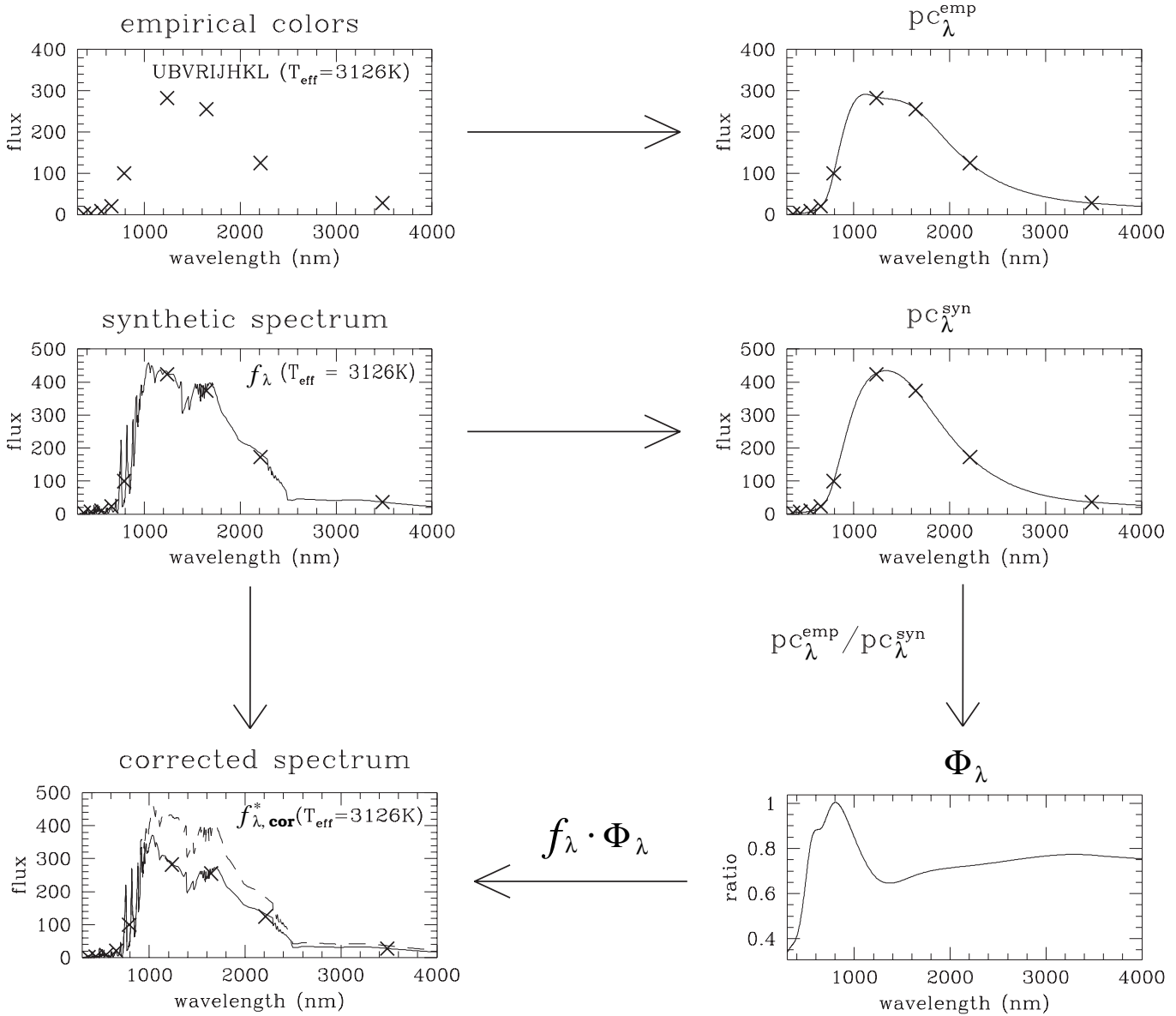}
\caption[]{Correction procedure. All fluxes are normalized to
        $f_{I}=100$.}
\label{f:corproc}
\end{figure*}

\begin{figure}[h]
\epsfxsize=9cm
\epsffile{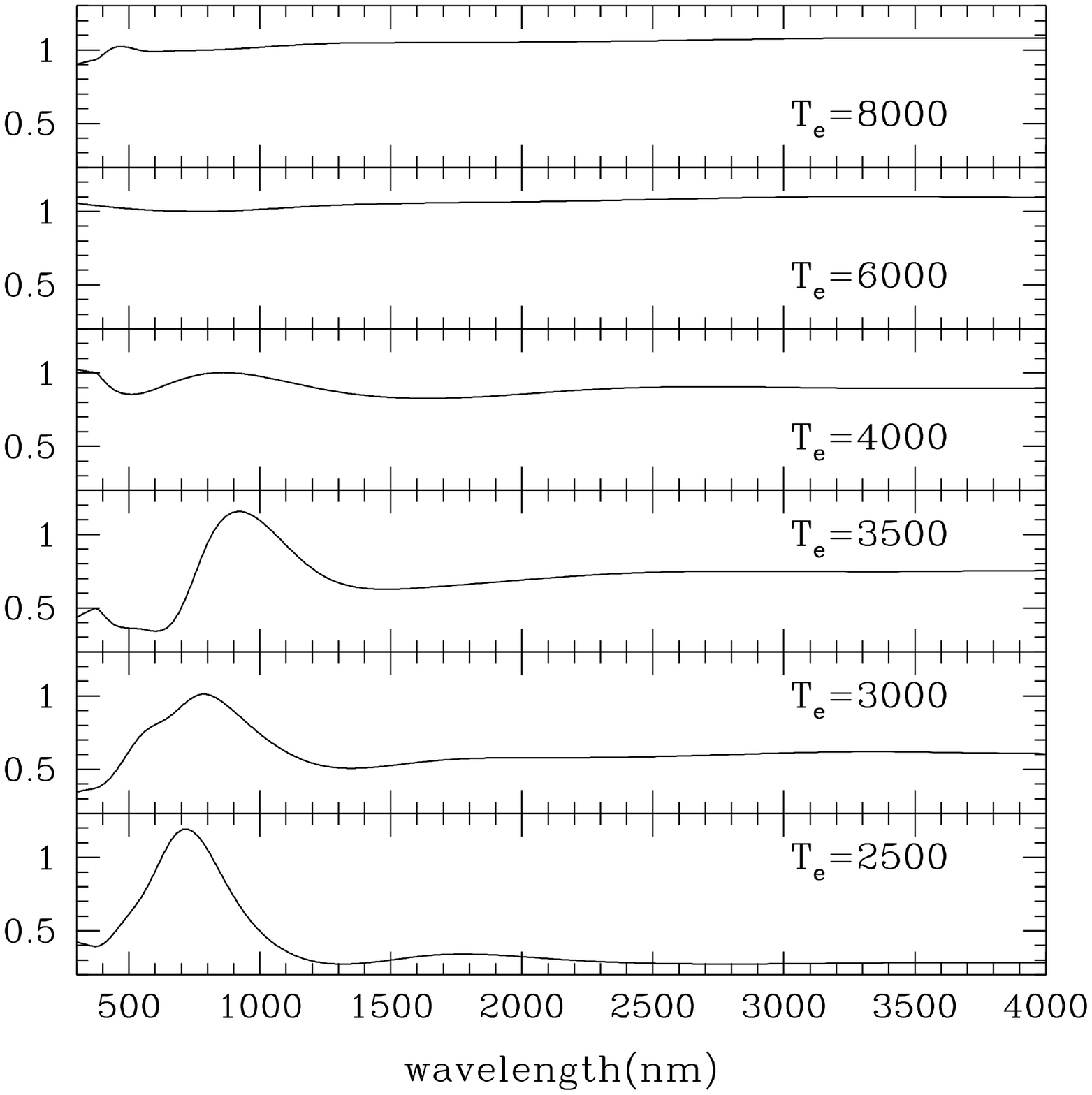}
\caption{Correction functions for a range of effective temperatures.}
\label{f:fcor}
\end{figure}

\item
Finally, the normalization   of  fluxes in the I--band,   $f_{i}=100$,
adopted   initially   for calculating  the   pseudo--continua must  be
cancelled    now   in order to     restore   the effective temperature
scale.    Thus,   each  corrected     spectrum    of  the     library,
$f_{\lambda,corr}^{*}$  is    scaled    by   a      constant   factor,
$\xi(T,logg,\chi)$,     to     give  the    final   corrected    spectrum
$f_{\lambda,corr}$:

\begin{equation}
f_{\lambda,corr}(T,logg,\chi)=\xi(T,logg,\chi) \cdot f_{\lambda,corr}^{*}(T,logg,\chi),
\end{equation}

where

\begin{equation}
\xi(T,logg,\chi)=\frac{\sigma T^{4}}
     {\int f_{\lambda}(T,logg,\chi) \cdot \Phi_{\lambda}(T) \cdot d\lambda}   
\end{equation}

assures that   the  emergent integral   flux  of  the  final corrected
spectrum conforms to  the  definition  of the effective    temperature
(eqn. (1)).

\end{enumerate}

\begin{figure*}
\epsfxsize=18cm
\epsffile{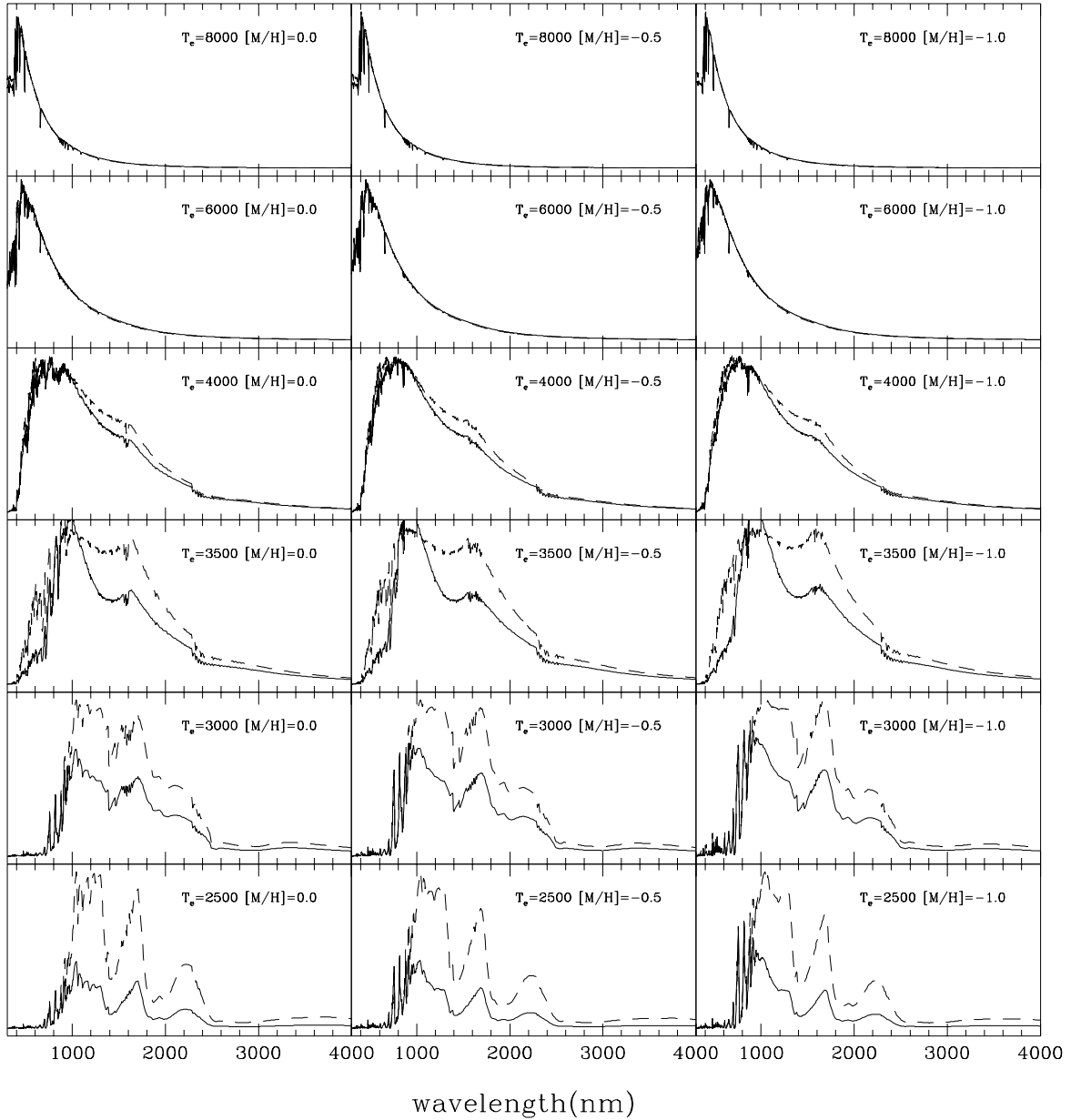}
\caption[]{Normalized corrected (solid lines) and original (dashed lines)
        library   spectra   for ranges in   effective  temperature and
        metallicity and  covering wavelengths from the photometric U--
        through K--passbands.  Top  panels:  K--library  dwarf models;
        middle panels:    K--library  giant  models;  bottom   panels:
        B+F--library giant models.}
\label{f:sp_spcor} 
\end{figure*}

\begin{figure*}
\epsfxsize=18cm
\epsffile{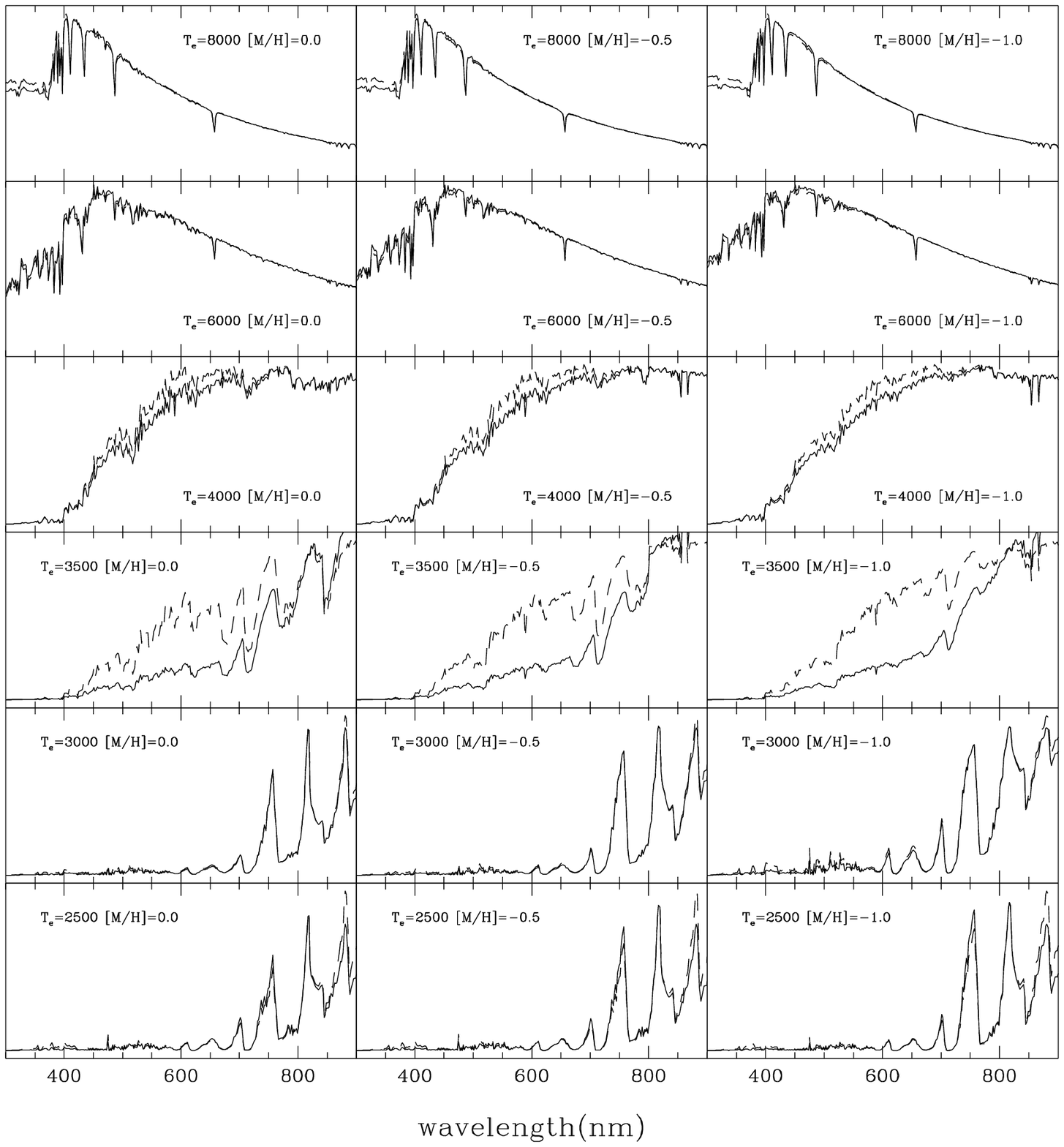}
\caption[]{Same as Fig. 12, but for the visible--near infrared wavelength
        ranges only.}
\label{f:sp_spcorvis}
\end{figure*}

The  final corrected   spectra  are  thus  in a   format which  allows
immediate applications  in population evolutionary  synthesis.  For  a
stellar model of  given mass, metallicity, and  age, the radius $R$ is
determined   by  calculations of  its     evolutionary  track in   the
theoretical HR diagram, and the total emergent flux at each wavelength
can hence be obtained from the present library spectra via

\begin{equation}
F_{\lambda} (T,logg,\chi)= 4 \pi R^2 \cdot f_{\lambda,corr}(T,logg,\chi).
\end{equation}

\section{Results: the corrected library spectra}

We now discuss the properties of the  new library spectra which result
from  the correction algorithm developed   above, and which are  most
important in the context of population and evolutionary synthesis.

\subsection{${\rm T_{eff}}$--color relations}

Fig.  14   illustrates the $T_{eff}$--color relations   obtained after
correction   of   the giant   sequence    spectra from   the   K-- and
B+F--libraries. Comparison    with the corresponding   Fig.  6 for the
uncorrected spectra shows  that the original differences which existed
both   between overlapping spectra of    the two libraries {\em and\/}
between    the synthetic and   empirical  relations have indeed almost
entirely been   eliminated.   While    remaining  differences  between
libraries  are negligible,  those  between  theoretical  and empirical
relations are below 0.1 mag.

\begin{figure*}
\epsfxsize=18cm
\epsffile{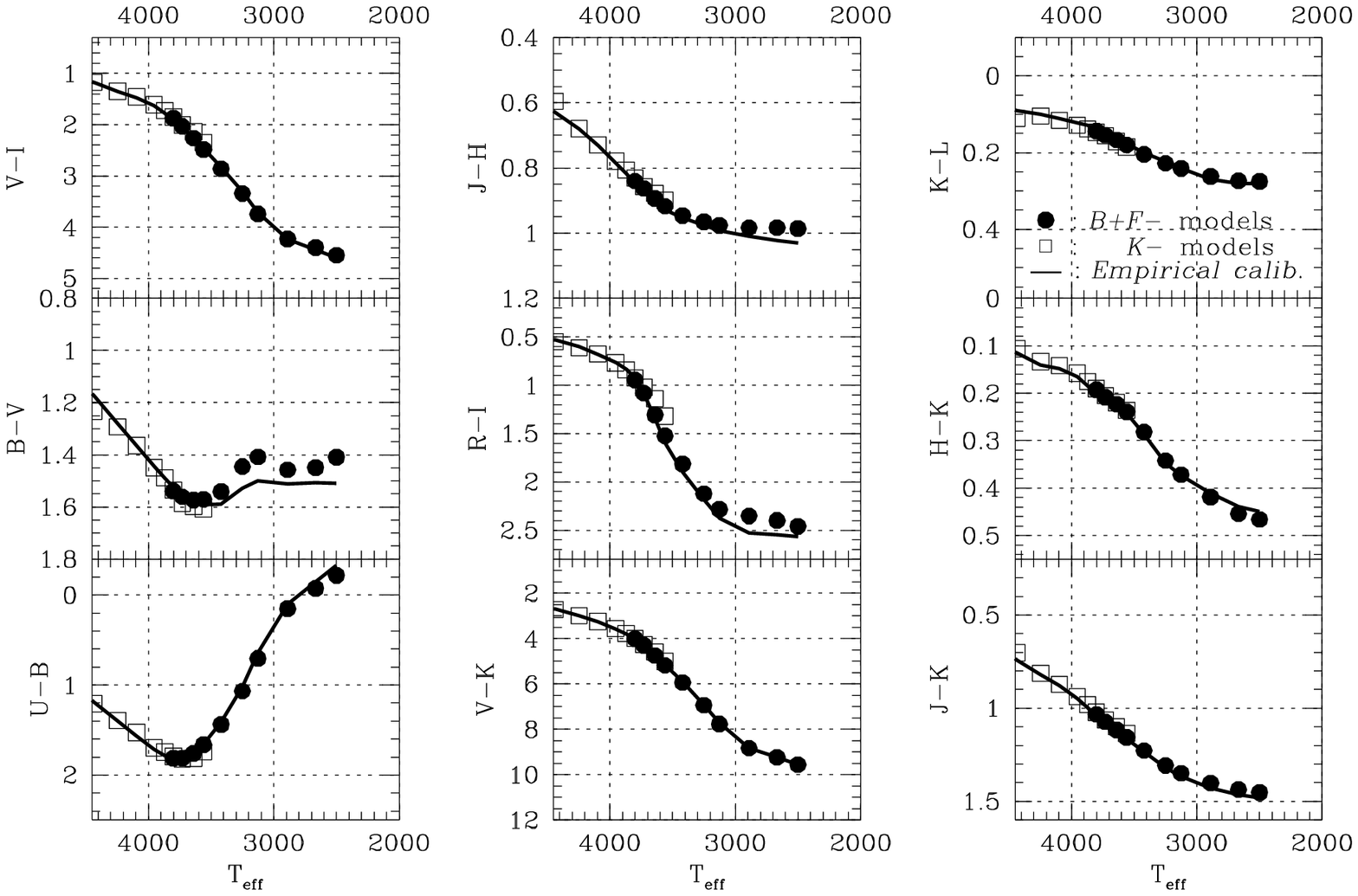}
\caption[]{Empirical color--effective temperature calibrations for
        solar--metallicity red  giant stars (solid lines, according to
        Table 2)  compared to  the corresponding theoretical relations
        calculated  from {\em  corrected\/} synthetic library  spectra
        (symbols, according to key in insert). Compare with Fig. 6.}
\label{f:gicol_cor}
\end{figure*}

Fig. 15 illustrates  similar  results for  the  main sequence.  Again,
comparison with the corresponding  Fig. 7 before correction shows that
the    present    calibration   algorithm     provides     theoretical
color--temperature relations  which  are in  almost perfect  agreement
with the empirical data.

\begin{figure*}
\epsfxsize=18cm
\epsffile{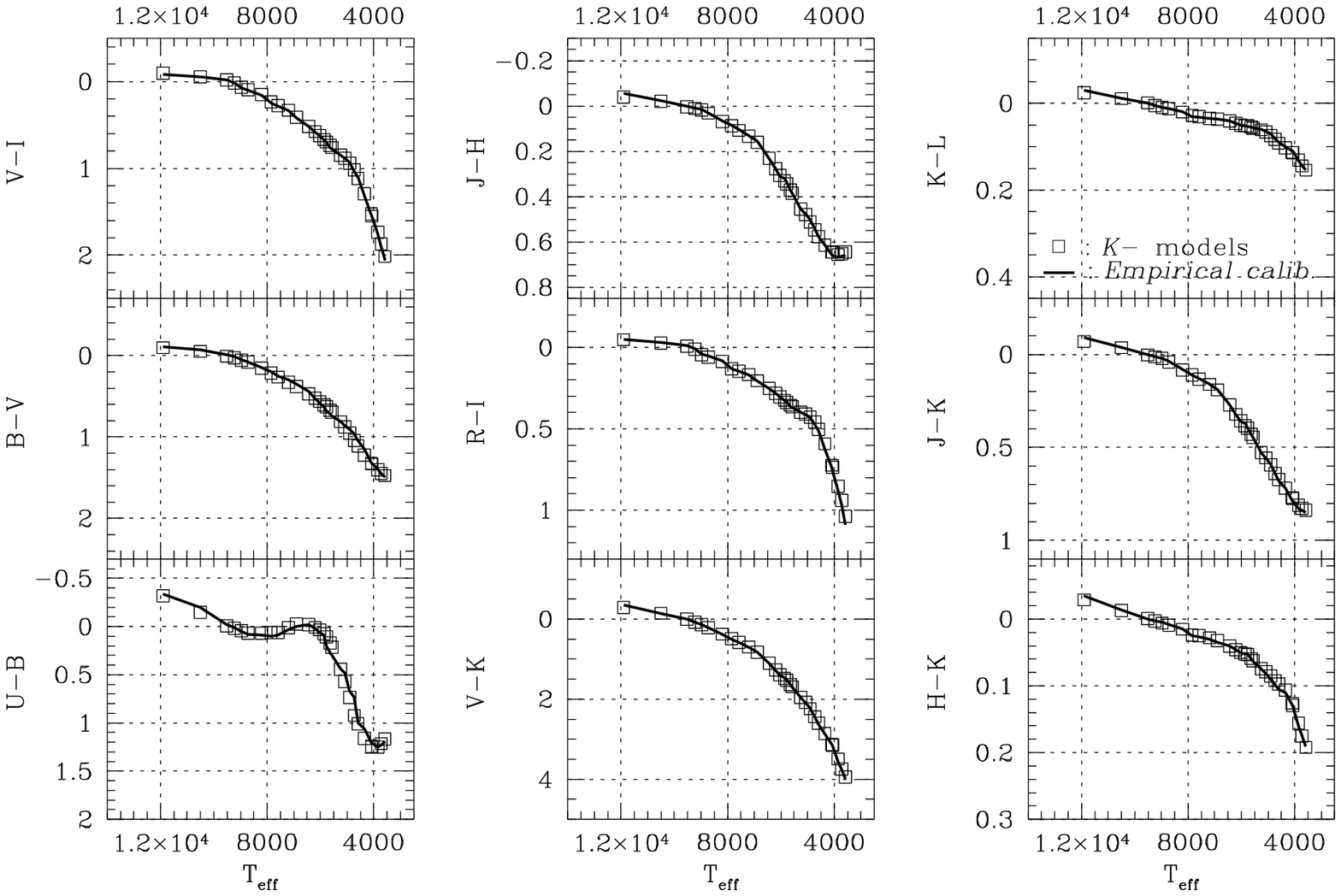} 
\caption[]{Empirical color--effective temperature calibrations for
        solar--metallicity  dwarf stars   (solid lines,  see  text for
        sources)  compared  to the corresponding theoretical relations
        calculated from {\em corrected\/} synthetic K--library spectra
        (symbols). Compare with Fig. 7.}
\label{f:dwcol_cor}
\end{figure*}

Thus at this point,   we can say  that  for solar abundances,  the new
library provides purely synthetic giant and dwarf star spectra that in
general fit empirical color--temperature calibrations to within better
than 0.1 mag over significant  ranges of wavelengths and temperatures,
and even  to  within a few hundredths  of  a magnitude for the  hotter
temperatures, $T_{eff}\geq4000K$.

\subsection{Bolometric corrections}

Bolometric corrections, $BC_{V}$,  are  indispensable  for the  direct
conversion of the theoretical HR diagram, $M_{bol}(T_{eff})$, into the
observational color--absolute magnitude diagram, $M_{V}(B-V)$:

\begin{equation}
BC_{V}=M_{bol}-M_{V} + {\rm constant},
\end{equation}

\noindent where the bolometric magnitude,

\begin{equation}
\begin{array}{rcl}
M_{bol} & = & -2.5 log \displaystyle \int_{0}^{\infty}f_{\lambda,corr} \, d\lambda \\
        &   &       \\
        & = & -2.5 log (\sigma T_{eff}^{4}/\pi), 
\end{array}
\end{equation}

provides the direct  link  to the  effective temperature  (scale).  Of
course,  bolometric corrections   applying to  any  other  (arbitrary)
passbands      are     then   consistently     calculated         from
$BC_{i}=BC_{V}+(M_{V}-M_{i})$,  where   the   color  $M_{V}-M_{i}$  is
synthesized from the corrected library spectra.

Fig.  16 provides  a  representative plot of   bolometric corrections,
$BC_{V}$,  for solar--abundance  dwarf  model  spectra.  The arbitrary
constant in Eq.   11 has  been  defined in  order  to fix to  zero the
smallest bolometric correction (Buser  \&  Kurucz 1978) found  for the
non-corrected    models,    which gives  $BC_{\odot}=BC_{(5577,  4.44,
0.0)}=-0.190$.  Comparison  with  the  empirical calibration given  by
Schmidt-Kaler   (1982)  demonstrates   that  the present   correction
algorithm is  reliable  in this  respect,  too: predictions everywhere
agree  with the empirical data  to within $\sim  0.05mag$  -- which is
excellent.  Similar tests for the giant models  also indicate that the
correction procedure  provides  theoretical  bolometric corrections in
better agreement  with the   observations.    These results   will  be
discussed in a subsequent paper based on a more systematic application
to multicolor data for  cluster and field  stars (Lejeune {\em et al.}
1997).

\begin{figure}
\epsfxsize=9cm
\epsffile{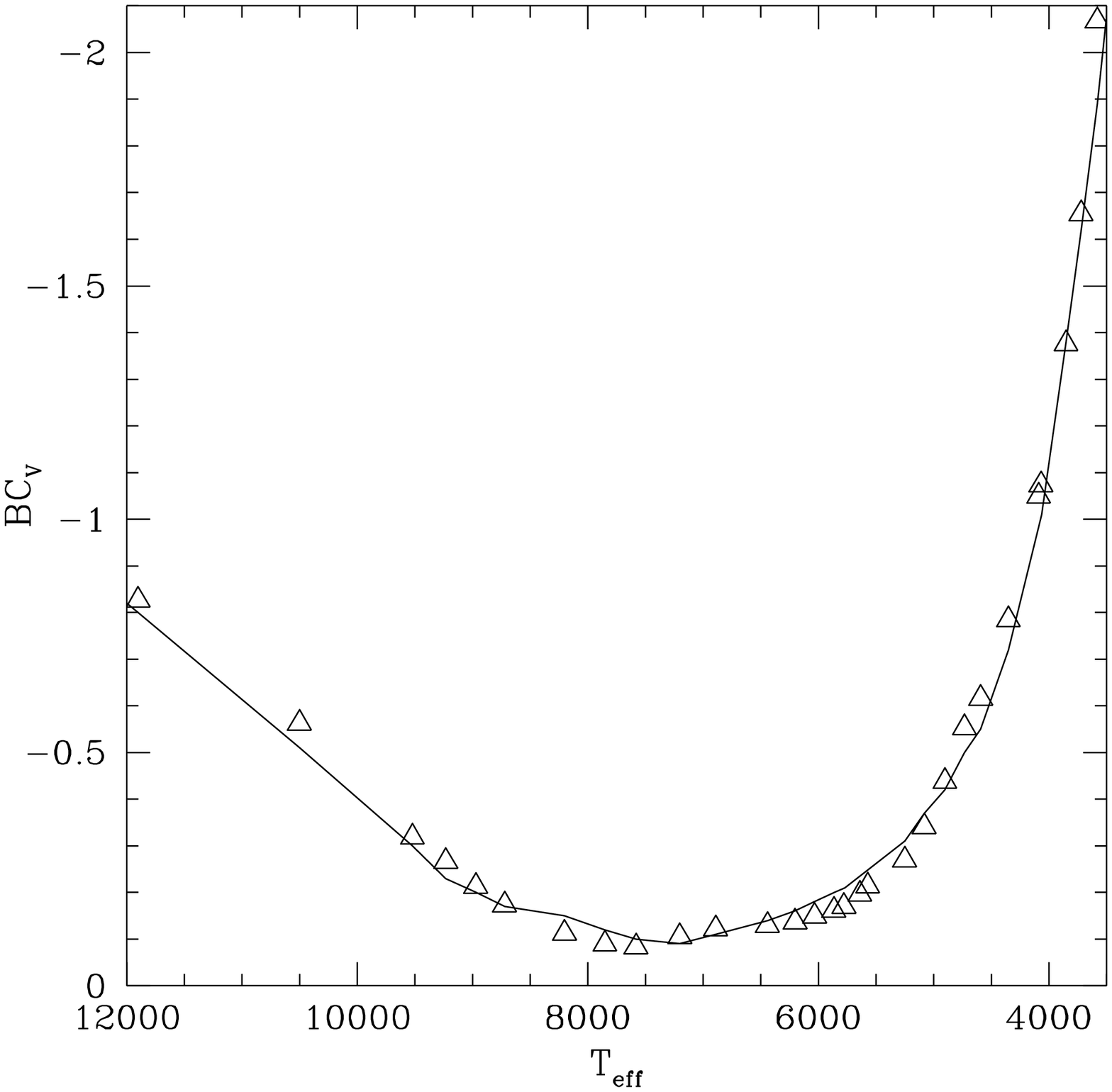}
\caption[]{Bolometric correction, $BC_{V}$, as a function of effective
        temperature,     for        {\em   corrected\/}     K--library
        solar--metallicity  dwarf  model   spectra. The  solid    line
        represents  the empirical calibration  given by  Schmidt-Kaler
        (1982).}
\label{f:bc_v}
\end{figure}

\subsection{Grid of differential colors}

Since comprehensive empirical     calibration  data have  only    been
available for the full temperature sequences of solar--abundance giant
and dwarf stars, {\em   direct\/} calibration of the  present  library
spectra using these data has, by  necessity, also been limited to {\em
solar--abundance\/} models.    However, because  one of  the principal
purposes of the  present work has been  to  make available theoretical
flux spectra  covering a  wide {\em  range  in metallicities\/}, it is
important that the  present calibration for solar--abundance models be
propagated  consistently into    the  remaining  library  spectra  for
parameter values ranging outside those  represented by the calibration
sequences.   We thus have designed  our correction algorithm in such a
way  as to {\em preserve,  at each temperature, the monochromatic flux
ratios between the  original spectra for different metallicities [M/H]
and/or surface gravities  log g\/}.  Justification of  this  procedure
comes  from the fact  that, if  used differentially,  most modern {\em
grids  of   model--atmosphere  spectra\/}  come  close  to reproducing
observed  stellar properties with  relatively high systematic accuracy
over wide  ranges in physical  parameters (e.g., Buser \& Kurucz 1992,
Lejeune \& Buser 1996).

In order to check the   extent to which preservation of  monochromatic
fluxes propagates into the  broad--band colors, we have calculated the
differential colors due  to metallicity differences between models  of
the same effective temperature and surface gravity:

\begin{equation}
\Delta(c_{j,[M/H]})=c_{j,[M/H]}-c_{j,[M/H]=0}, j=1,...,8.
\end{equation}

We can  then  calculate  the  residual  color differences between  the
corrected and the original grids:

\begin{equation}
\delta(\Delta(c_{j,[M/H]}))=
	\Delta(c_{j,[M/H]}^{corr})-\Delta(c_{j,[M/H]}^{orig}).
\end{equation}

Results are  presented in Figs. 17  and 18  for the coolest K--library
models $(3500K \leq  T_{eff}  \leq 5000K)$  and for  the  B+F--library
models      for M  giants    $(2500K   \leq    T_{eff}  \leq  3750K)$,
respectively. Residuals are plotted as a function of the model number,
which  increases with both surface  gravity and effective temperature,
as  given in the calibration  sequences. The different lines represent
different metallicities, $-3.0 \leq  [M/H] \leq +1.0$, as explained in
the captions.

\begin{figure}
\epsfxsize=9cm
\epsffile{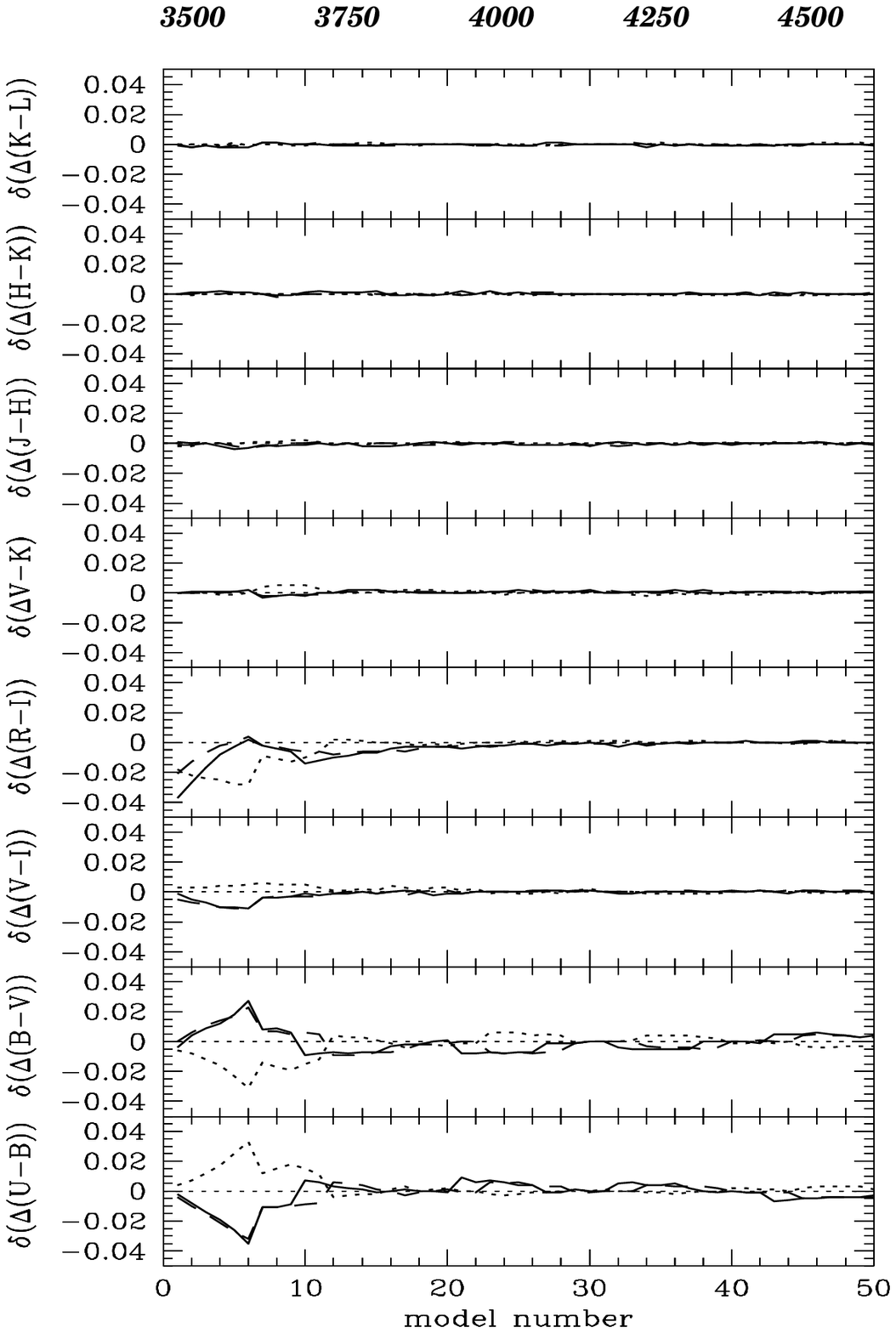}
\caption []{Residuals between differential colors obtained from original
        and corrected library spectra for models of the same effective
        temperatures       and  surface   gravities   but    different
        metallicities. The  abscissa represents the monotonic increase
        in temperature (as  indicated along the  top of the panel) and
        surface gravity in a similar way as Table 2. Residuals are for
        the  coolest models from  the K--library  $(3500K \sim 5000K)$
        and  for metallicities $[M/H]=-3.0$ (solid line), $[M/H]=-1.0$
        (long-dashed line), and $[M/H]=+1.0$ (dotted line).}
\label{f:diff_effect.K95}
\end{figure}
\begin{figure}
\epsfxsize=9cm
\epsffile{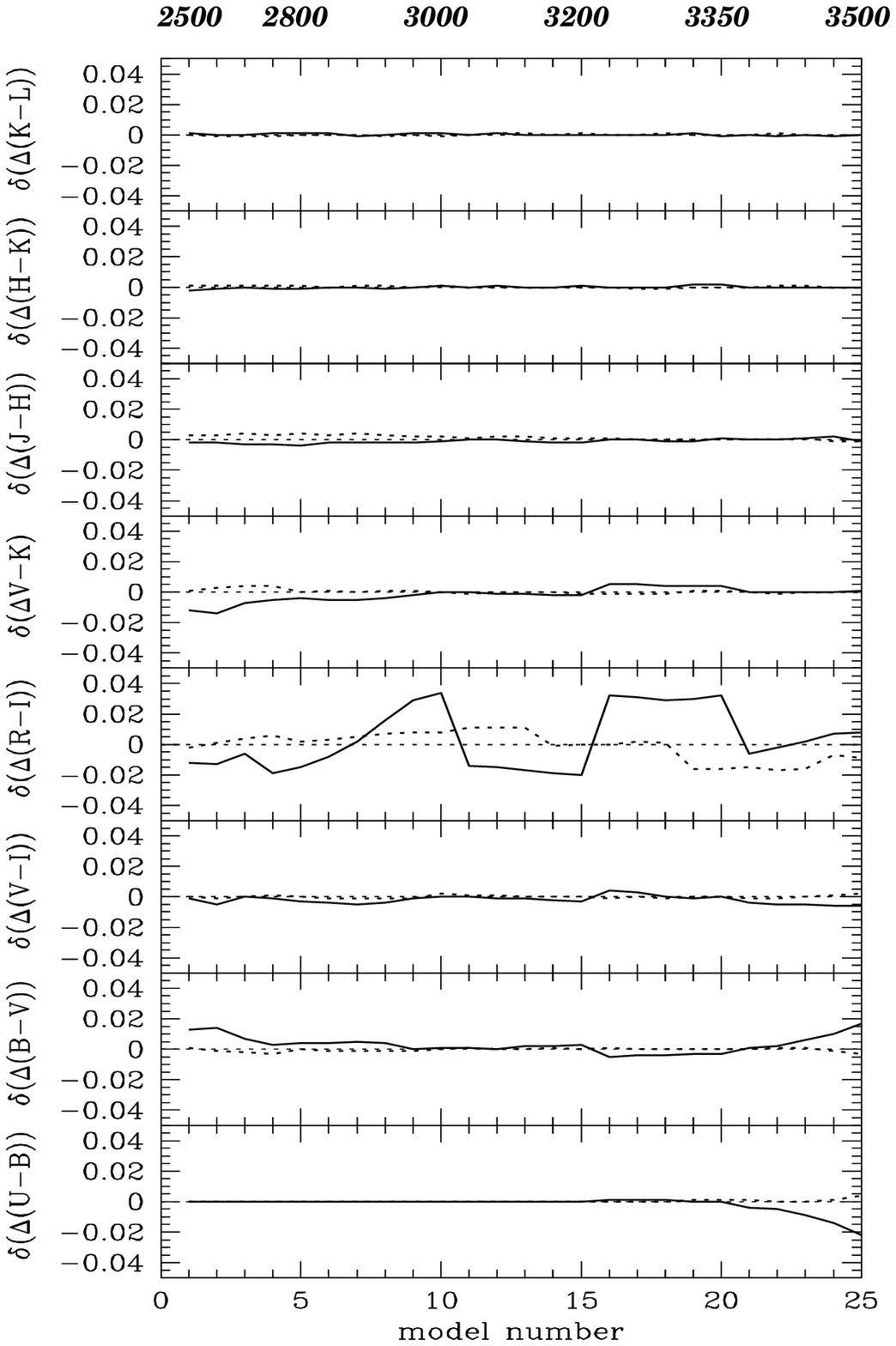}
\caption []{Same as Fig. 17, but for the M--giant models from the
        B+K--library $(2500K \sim 3750K)$ and for metallicities
        $[M/H]=-1.0$ (solid line) and $[M/H]=+0.5$ (dotted line). The zero
        residuals for $U-B$ are due to the fact that in this wavelength range,
        B+K--library spectra are available for solar metallicity only.}
\label{f:diff_effect.FB}
\end{figure}

The most  important  conclusion is  that, in  general,  the correction
algorithm does not alter  the  original differential grid   properties
significantly for  most colors and most  temperatures --  in fact, the
residuals  are  smaller than only  a  few hundredths   of a magnitude.
Typically, the  largest    residuals   are found   for   the   coolest
temperatures $(T_{eff}  \leq   3800K)$  and   the shortest--wavelength
colors, UBVRI,  where  the correction functions  of Fig.   11 show the
largest variations not only  between the different passbands, but also
{\em within  the individual passbands\/}.     This changes their  {\em
effective wavelengths\/} and, hence, the  baselines defining the color
scales (cf. Buser  1978).  Since this  effect tends  to grow with  the
width of the  passband, it is  mainly the coincidence of large changes
in both amplitudes {\em and\/} slopes of the correction functions with
the wide--winged R--band which causes residuals  for the R-I colors to
be relatively large in Fig. 18.

Calculations     of color  effects induced   by   {\em surface gravity
changes\/}  lead to similar results.  This corroborates our conclusion
that  the present  correction  algorithm indeed  provides a new  model
spectra  library  which  essentially  incorporates,  to  within useful
accuracy for the purpose,  the currently best knowledge of fundamental
stellar properties: a full--range  color--calibration in terms of {\em
empirical\/}   effective   temperatures  at   solar  abundances (where
comprehensive calibration data exist) {\bf and\/} a systematic grid of
differential colors   predicted by   the original  {\em theoretical\/}
model--atmosphere calculations  for the full  ranges of  metallicities
and surface gravities  (where empirical data are  still too scarce  to
allow comprehensive grid calibration).

Of course, we are aware  that the present correction algorithm becomes
increasingly inadequate with  the  complexity of  the stellar  spectra
growing with decreasing temperature  and/or increasing surface gravity
and metallicity. For  example,   because under these   conditions  the
highly {\em nonlinear\/} effects  of blanketing due to line saturation
and crowding and broad--band molecular absorption tend to dominate the
behavior  of stellar colors,  particularly  at shorter (i.e., visible)
wavelengths, even the  corresponding differential colors cannot either
be recovered  in a physically   consistent manner by a  simple  linear
model such as  the present. However, the limits  of this approach will
be further explored in Paper II,  where the calibration of theoretical
spectra    for  M--dwarfs will    be  attempted   by  introducing  the
conservation of original differential colors of grid spectra {\em as a
constraint\/} imposed to the correction algorithm.

\section{Organization of the library}

The corrected spectra have  finally   been composed into  the  unified
library shown in Fig. 19. In this library, model spectra are given for
a parameter grid which  is uniformly sampled in  $T_{eff}$, log g, and
[M/H], each  spectrum being  available for  the same wavelength  grid,
(${\rm  91 \AA  \leq   \lambda \leq  1,600,000  \AA}$), with  a medium
resolution of ~ 10--20 \AA.

\begin{figure}
\epsfxsize=9cm
\epsffile{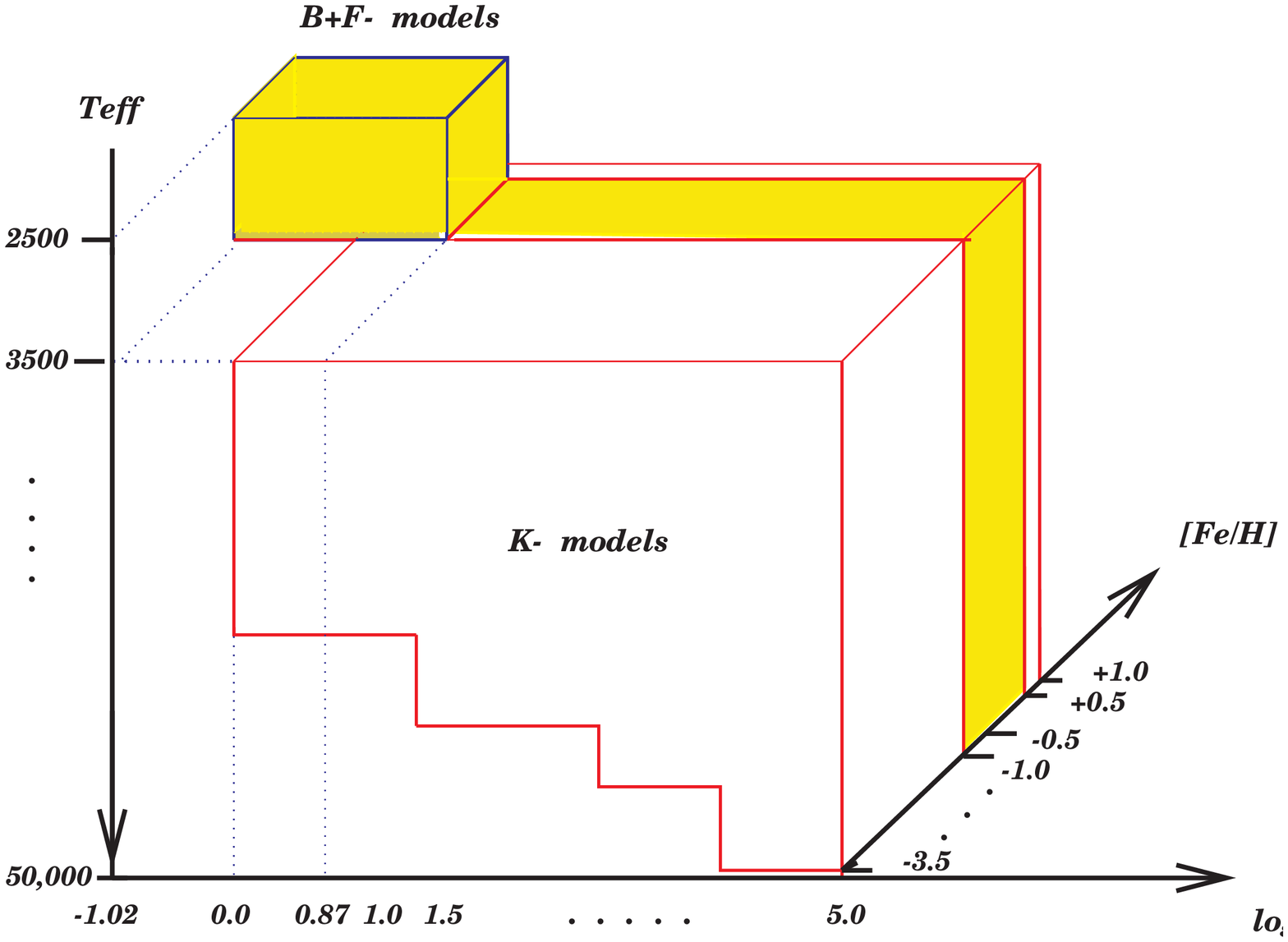}
\caption []{The $(\rm T_{eff}, log g, [M/H])$ parameter space covered by
        the present unified library. Solid lines delineate approximate
        boundaries of available models. (B+F)--models for M--giants are
        confined to the metallicity range $-1.0 \leq [M/H] \leq +0.5$
        (grey part).}
\label{f:bcl96.3D}
\end{figure}

The main  body  consists of the  K--library, which  provides  the most
extensive   coverage of  parameter  space  (cf.   Table  1) down    to
$T_{eff}=3500K$, and which  has  therefore been  fully  implemented to
this limit. The extensions  to lower temperatures, $3500K \geq T_{eff}
\geq 2500K$, and associated lower  surface gravities, $0.87 \geq log g
\geq   -1.02$,  are  provided   by  the   M--giant  spectra from   the
B+F--library,   which, however, covers   only  the limited metallicity
range $+0.5 \geq [M/H] \geq -1.0$.

This unified library is  available (in  electronic  form) at  the CDS,
Strasbourg, France,   where it can   be   obtained in  either  of  two
versions, with  the  {\em corrected\/}  {\bf or\/}  the  {\em original
uncorrected\/} spectra from the K-- and B+F--libraries.

\section{Discussion and conclusion}

Although astronomers  have   for a  long time agreed   that a uniform,
complete, and realistic stellar library is urgently needed, it must be
emphasized that this goal has remained too ambitious to be achieved in
a single  concerted effort through  the  present  epoch. We have  thus
attempted  to  proceed in well--defined    steps, with priorities  set
according to the  availability of  basic  data and following the  most
obvious scientific questions that  would likely become more tractable,
or  even answerable.    Therefore,   we  briefly review  the   present
achievement  to clarify  its status in  the  ongoing process toward  a
future  {\em standard  library\/} of  theoretical stellar spectra  for
photometric evolutionary synthesis.

\begin{enumerate}
\item {\bf Completeness.}
The  unification  of the  massive  K--library  spectra  with those for
high-- luminosity M--star models (the B+F--library) is most important,
because even as a  small minority of the  number population of a given
stellar system, the late--type   giants and supergiants may  provide a
large   fraction  of this  system's   integrated light at  visible and
infrared wavelengths.  This fact was  recognized early  on (e.g., Baum
1959), and eventually also   co--motivated the effort leading to  the
existence of the B--library used in  this work (Bessell {\em et al.\/}
1988).

While the K--,  B--,    and F--libraries have   been  used  to  remedy
incompleteness (in either wavelength  or parameter coverage,  or both)
of available {\em observed\/} stellar  libraries before (e.g., Worthey
1992, 1994; Fioc  \& Rocca--Volmerange 1996), our  first goal here has
been to join  them as a  purely {\em theoretical\/} library, providing
the main advantages of physical homogeneity and definition in terms of
fundamental stellar parameters -- which allows direct use with stellar
evolutionary calculations.

But   even  so, the  present  library remains   incomplete  in several
respects. First, stellar evolution  calculations (e.g., Green  {\em et
al.\/} 1987)  predict  that high--luminosity  stars with  temperatures
near or below $3500K$ may also exist at low metallicities, $[M/H] \sim
-1.0$,  and  their flux    contributions at visible--near  ultraviolet
wavelengths (where  metallicity produces significant effects)  may not
quite  be   negligible  in  the   integrated  light   of  old  stellar
populations.   Therefore, in order   to  provide  the  fuller coverage
required for an adequate study  of this metallicity--sensitive domain,
new calculations of B--library spectra extending  the original data to
both  shorter  wavelengths  $(\lambda\lambda \geq   320nm)$ and  lower
abundances $([M/H] \geq -2)$ (Buser {\em et al.\/} 1997) will replace
the current hybrid B+F spectra and make the  next library version more
homogeneous.

Secondly, even though  the low--temperature, low--luminosity \/ M--dwarf
stars  do not  contribute  significantly to the integrated  bolometric
flux, they are not  negligible in the determination of mass--to--light
ratios in stellar populations.  Thus, a suitable grid of (theoretical)
M--dwarf spectra calculated  by Allard \&  Hauschildt (1995)  is being
subjected  to a  similar calibration process  (Paper II)  and  will be
implemented  in  the present library as  an  important step toward the
intended {\em standard stellar library\/}.

Finally, the libraries of synthetic spectra  for hot O-- and WR--stars
which were recently  calculated by Schmutz {\em  et al.\/} (1992) will
allow  us  to  extend  the calibration   algorithm  to ultraviolet IUE
colors, where such stars radiate most of their light.
\\
\item {\bf Realism.}
In view of its major intended  application -- photometric evolutionary
synthesis  --,  the  {\em minimum  requirement\/}  that we  insist the
theoretical  library {\em must\/} satisfy,  is to provide stellar flux
spectra   having  (synthetic) colors  which   are  systematically {\em
consistent with  calibrations derived from  observations\/}. How  else
could we hope to learn the physics of distant stellar populations from
their integrated colors, unless the basic building blocks -- i.e., the
library spectra used in the synthesis calculations  -- can be taken as
adequate representations  of   the better--known fundamental   stellar
properties, such as their color--temperature relations?

Because the original  library spectra do  not  meet the  above minimum
requirement (Sect. 3), we  have developed an algorithm for calibrating
existing theoretical  spectra against     empirical color--temperature
relations   (Sect.    4).  Because  comprehensive  empirical  data are
unavailable for  large segments of the  parameter space covered by the
theoretical  library, direct calibration  can be effected only for the
major  sequences of solar--abundance models   (Sect.  5).  However, we
have   also shown that  the   present  algorithm provides the  desired
broad--band (or pseudo--continuum)  color calibration  {\em without\/}
destroying   the  original   relative    monochromatic fluxes  between
arbitrary model grid spectra and solar--abundance calibration sequence
spectra  of the   same  effective temperature.    This conservation of
original grid properties   also   propagates with  useful   systematic
accuracy even through most {\em  differential broad--band colors\/} of
the corrected library spectra.  Thus, to  the extent that differential
broad--band colors of original  library spectra were  previously shown
to be consistent with spectroscopic or other empirical calibrations of
the  UBV--,  RGU--,  and   Washington ultraviolet-excess--metallicity
relations (Buser  \& Fenkart  1990, Buser \&  Kurucz 1992,  Lejeune \&
Buser 1996), the {\em  corrected} library spectra are still consistent
with the same calibrations.
\\
\item{\bf Library development.}
At this   point, we feel   that some of  the more  important intrinsic
properties required of  the   future  {\em standard  library\/}   have
already been established.  Of course,  many more consistency tests and
calibrations  will now be needed that  can, however, only be performed
for {\em  local\/} volumes of the full  parameter space covered by the
new  library. For example,  we shall  use  libraries of  observed flux
spectra  for individual field  and  cluster stars  to better assess --
and/or improve -- the performance   of the present library version  in
the     non--solar--abundance       and      non--visual    wavelength
regimes. Eventually, we also expect significant guidance toward a more
{\em systematically\/} realistic   version of the  library from actual
evolutionary   synthesis calculations  of  the  integrated spectra and
colors of globular clusters (Lejeune 1997).

Last,  but not least, we  would like to  emphasize that, while we here
present the results  taylored according  to the  general needs  in the
field of photometric  evolutionary synthesis, the library construction
algorithm has been designed such  as to {\em allow flexible adaptation
to  alternative  calibration data as  well\/}.   As we shall ourselves
peruse this flexibility  to accommodate both  feed--back and new data,
the  reader, too,  is  invited  to define   his or  her own  preferred
calibration constraints and   have the  algorithm adapted  to  perform
accordingly.

\end{enumerate}

{\em Acknowledgements}. We are grateful to  Michael Scholz and Gustavo
Bruzual for providing vital input and critical discussions. Christophe
Pichon  is also  aknowledged  for  his precious   help  with the final
implementation of some of  the figures.  We wish  to thank warmly  the
referee for   his  helpful comments and   suggestions.  This  work was
supported by the Swiss National Science Foundation.

\end{document}